# MWF-MIMOSA for efficient simultaneous relaxometry and myelin water fraction mapping

Yuting Chen[1,2,3], Yohan Jun#[2,3], Hyeong-Geol Shin[4,5], Shizhuo Li[1,2,3], Shohei Fujita[2,3], Xingwang Yong[2,3], Jiye Kim[6], Jongho Lee[6], Gian Franco Piredda[7], Tom Hilbert[7-9], Aneri Bhatt[2,3], Susie Y. Huang[2,3], Huafeng Liu#[1,10], Huihui Ye[11], Shahin Nasr[2,3], Borjan Gagoski[3,12], Kwok-Shing Chan[2,3], and Berkin Bilgic[2,3,13]

[1]State Key Laboratory of Extreme Photonics and Instrumentation, College of Optical Science and Engineering, Zhejiang University, Hangzhou, China, [2]Athinoula A. Martinos Center for Biomedical Imaging, Massachusetts General Hospital, Boston, MA, United States, [3]Department of Radiology, Harvard Medical School, Boston, MA, United States, [4]Johns Hopkins University, Baltimore, MD, United States, [5]F.M. Kirby Research Center, Kennedy Krieger Research Institute, Baltimore, MD, United States, [6]Department of Electrical and Computer Engineering, Seoul National University, Seoul, Republic of Korea, [7]Swiss Innovation Hub, Siemens Healthineers International AG, Lausanne, Switzerland, [8]Department of Radiology, Lausanne University Hospital and University of Lausanne, Lausanne, Switzerland, [9]LTS5, École Polytechnique Fédérale de Lausanne (EPFL), Lausanne, Switzerland, [10]Key Laboratory for Biomedical Engineering of Ministry of Education, College of Biomedical Engineering and Instrument Science, Zhejiang University, Hangzhou, China, [11]School of Communication Engineering, Hangzhou Dianzi University, Hangzhou, China, [12]Fetal-Neonatal Neuroimaging & Developmental Science Center, Boston Children's Hospital, Boston, MA, United States, [13]Harvard/MIT Health Sciences and Technology, Massachusetts Institute of Technology, Cambridge, MA, United States

**Word count**: Approximately 7000 words

**Corresponding author:**

# Huafeng Liu and Yohan Jun are co-corresponding authors.

Huafeng Liu

Email: liuhf@zju.edu.cn

Address: State Key Laboratory of Extreme Photonics and Instrumentation, College of Optical Science and Engineering, Zhejiang University, Hangzhou 310027, China.

Yohan Jun

Email: yjun@mgh.harvard.edu

Address: Athinoula A. Martinos Center for Biomedical Imaging, Building 149, 13th Street, Rm 2301, Boston, MA 02129, USA.

**ORCID:**

Yuting Chen: 0009-0003-6232-1141; Yohan Jun: 0000-0003-4787-4760;

Berkin Bilgic: 0000-0002-5806-4897.

# Abstract

Quantitative magnetic resonance imaging (qMRI) provides improved sensitivity and specificity to tissue composition and pathological alterations compared with conventional contrast-weighted imaging. Among various qMRI biomarkers, myelin water imaging is of particular interest because myelin plays a central role in brain function and its alteration is closely associated with many neurological diseases. However, conventional myelin water fraction (MWF) mapping techniques are often limited by long scan times, low spatial resolution, reduced signal-to-noise ratio (SNR), and high specific absorption rate (SAR). Here, we propose MWF-MIMOSA for efficient simultaneous $T_1$, $T_2$, $T_2$* mapping, magnetic susceptibility source separation, and MWF estimation. To achieve this, multi-contrast and multi-slice zero-shot self-supervised learning (MZS-SSL) was used to jointly reconstruct whole-brain complex-valued images. To improve computational efficiency of the parameter estimation step, a multilayer perceptron (MLP) was trained within the GACELLE GPU-accelerated parameter estimation framework to circumvent the computationally intensive Bloch simulation process, resulting in a >100-fold computational speed-up in MWF estimation. Numerical simulations were performed to evaluate the accuracy and precision of MWF-MIMOSA, and in-vivo results further demonstrated its robustness. Comparison with existing myelin water imaging methods showed that MWF-MIMOSA is highly correlated with established approaches, while providing complementary quantitative parameter maps at higher spatial resolution and with shorter scan times. Notably, simultaneous multi-parametric mapping was achieved in 5 min at 1 mm isotropic resolution, and in 10 min at 0.7 mm isotropic resolution. These results demonstrate the potential of MWF-MIMOSA for fast, high-resolution simultaneous relaxometry and myelin water imaging.

**Funding information:**

This work was supported by research grants NIH R01 EB032378, R01 EB028797, P41 EB030006, U01 EB026996, UH3 EB034875, R21 AG082377, R21 EB036105, R01 EB034757, and supported in part by the National Key Research and Development Program of China (No. 2020AAA0109502), by the National Natural Science Foundation of China (No. U1809204, 61701436), by the Zhejiang Provincial Natural Science Foundation of China (No. LY22F010007), and by Zhejiang Provincial “Jianbing Lingyan + X” Science and Technology Program (No. 2025C01127), the NVIDIA Corporation for computing support. KC was supported by NWO/ZonMw under the Rubicon award number 04520232330012.

# 1. INTRODUCTION

Quantitative MRI (qMRI) offers a versatile framework for characterizing tissue microstructure by estimating biophysical parameters (Weiskopf *et al.*, 2021) such as relaxation times (Okubo *et al.*, 2017; Thaler *et al.*, 2017), magnetic susceptibility (Fushimi *et al.*, 2024; Ma *et al.*, 2025), and macromolecular content (Gozdas *et al.*, 2021). These parameters reflect distinct aspects of tissue composition, including water mobility (Jara *et al.*, 2022), iron concentration (Langkammer *et al.*, 2012; Filo *et al.*, 2023), and myelin integrity (Lee *et al.*, 2021). However, individual quantitative parameters are inherently confounded by competing co-localized influences, representing the aggregate influence of multiple biophysical sources, including water mobility, iron concentration, and myelin deposition (Wu *et al.*, 2017; Yao *et al.*, 2018; Rahmanzadeh *et al.*, 2022). Conventionally, acquiring these parameters requires multiple separate scans, resulting in prohibitively long exams, poor patient compliance, and inter-scan motion artifacts that compromise voxel-wise analysis, making translating into clinical practice or research studies challenging (Mancini *et al.*, 2020). Therefore, an integrated acquisition framework capable of simultaneously deriving all these parameters in a single scan with high efficiency is highly desirable.

Multi-parametric mapping acquires a diverse set of complementary parameters to disentangle complex physiological processes. Through dedicated sequence design, multi-parametric MRI enables efficient encoding of multiple tissue properties in a single acquisition, allowing simultaneous quantification of multiple parameters while reducing overall acquisition time. Representative examples include magnetic resonance fingerprinting (MRF) (Ma *et al.*, 2013), echo planar time-resolved imaging (EPTI) (Wang *et al.*, 2022), MR multitasking (Christodoulou *et al.*, 2018), and 3D-quantification using an interleaved Look-Locker acquisition sequence with $T_2$ preparation pulse (3D-QALAS) (Kvernby *et al.*, 2014). In particular, Multi-Parametric Imaging Using Multiple-Echoes With Optimized Simultaneous Acquisition (MIMOSA) (Chen *et al.*, 2026) has been developed to enable whole-brain simultaneous $T_1$, $T_2$, $T_2^*$, QSM, para- and dia-magnetic susceptibility mapping with high accuracy and scan-rescan repeatability.

Despite acquiring multiple parameters simultaneously with high efficiency and spatial resolution, these multi-parametric mapping methods typically yield only single-compartment information. The rapidly decaying signals of myelin water are averaged with the dominant intra- and extracellular spaces. Myelin plays a critical role in ensuring rapid neural conduction and axonal support (Nave, 2010), and abnormal myelination is implicated in a wide range of neurological and psychiatric disorders, including multiple sclerosis (Choi *et al.*, 2019) and leukodystrophy (Lancione *et al.*, 2024; Yska *et al.*, 2025). Myelin water imaging (MWI) probes the proportion of water trapped between myelin bilayers to the total water volume (i.e., myelin water fraction, MWF) and has emerged as a promising quantitative imaging marker of myelin content (Laule *et al.*, 2006) beyond conventional qualitative measures such as $T_1$w/$T_2$w ratio imaging (Glasser and Van Essen, 2011), enabling investigations of myelin-related changes in brain development, aging, and neurological disorders (Faizy *et al.*, 2020; Rahmanzadeh *et al.*, 2022).

Conventional MWF mapping relies on a single relaxation contrast mechanism (Piredda, Hilbert, Thiran, *et al.*, 2021), such as $T_1$ (Oh *et al.*, 2013), $T_2$ (MacKay *et al.*, 1994), or $T_2^*$ (van Gelderen *et al.*, 2012; Nam *et al.*, 2015; Shin *et al.*, 2019), to quantify multiple tissue compartments. Multi-echo spin echo (MESE) (MacKay *et al.*, 1994) and multi-echo gradient and spin echo (GRASE) (Prasloski *et al.*, 2012; Piredda, Hilbert, Canales-Rodríguez, *et al.*, 2021) effectively capture the short $T_2$ (≤ 25 ms at 3T) signal of myelin water, but suffer from long scan times, low acquisition efficiency, and high specific absorption rate (SAR). Multi-echo gradient echo (MGRE) approaches leverage higher acquisition efficiency and low SAR by exploiting both compartmental $T_2^*$ decay and myelin-related frequency shifts differences (van Gelderen *et al.*, 2012; Nam *et al.*, 2015). However, MGRE-based methods are susceptible to $T_1$-weighting biases (Chan and Marques, 2020), particularly when using short repetition times (TR) with 3D encoding. Moreover, all these methods suffer from inherently ill-conditioned model fitting, demanding high signal-to-noise ratio (SNR) data and thereby limiting spatial resolution.

Recent advances have sought to achieve high-resolution MWF mapping to provide comprehensive brain tissue characterization with whole brain coverage (Wu *et al.*, 2017; Pu *et al.*, 2020). Multi-contrast approaches, such as multi-compartment relaxometry for MWI (MCR-MWI) (Chan and Marques, 2020; Chan, Chamberland and Marques, 2023), mcDESPOT (Deoni, Rutt and Peters, 2003; Bouhrara and Spencer, 2017), and VFA-EPTI (Dong *et al.*, 2021), acquire multiple flip angles to encode $T_1$ relaxation information along with $T_2/T_2^*$ information to improve the conditioning of MWF estimation. Although these methods can achieve higher resolution, the acquisition times remain relatively long, which is suboptimal for clinical adoption. Magnetic resonance fingerprinting (MRF) (Ma *et al.*, 2013) has also been adapted for multi-parametric MWI through 2D and 3D variants (e.g., PV-MRF (Chen *et al.*, 2019; Deshmane *et al.*, 2019) and 3D MRF-MWF (Lancione *et al.*, 2024). Sequences like MWF-QALAS (Fujita *et al.*, 2025) use short $T_2$-preparation pulses to capture short $T_2$ components efficiently. Despite these advances, most MRF-based methods estimate MWF using a partial-volume model (Deshmane *et al.*, 2019) in which the relaxation parameters of each compartment are fixed and assumed to be known a priori. This assumption can introduce significant bias, particularly in pathological conditions where the underlying tissue relaxation properties deviate from assumed baseline values. Beyond pathology, physical confounds can also violate such assumptions. In particular, $T_2^*$ is strongly dependent on fiber orientation relative to the main magnetic field (Li *et al.*, 2009; Lee *et al.*, 2011).

In this study, we propose MWF-MIMOSA, an integrated multi-parametric framework for the simultaneous mapping of $T_1$, $T_2$, $T_2^*$, QSM, susceptibility source separation, and MWF. The incorporation of an MGRE module into the sequence timing provides both $T_2^*$ decay and frequency shift information, which, when combined with preparation modules that confer $T_1$ and $T_2$ sensitivity to myelin water, enables the joint estimation of compartment-specific relaxation properties. This integrated strategy reduces reliance on fixed relaxation parameter assumptions and improves the robustness and physiological specificity of MWF estimation. The main contributions of our work are as follows:

- We proposed MWF-MIMOSA by extending the MIMOSA sequence to improve sensitivity to myelin-related signals and to enable multi-compartment microstructural modeling.

- We developed a three-compartment MWF estimation model based on Bloch simulations, without making assumptions about the relaxation parameters of the MW compartment to avoid potential biases.
- We implemented a multilayer perceptron (MLP) within the GPU-accelerated GACELLE framework (Chan *et al.*, 2025) to surrogate the time-consuming Bloch simulation process, achieving > 100-fold computational speed-up in MWF estimation.
- We performed simulations to systematically evaluate the accuracy and precision of the framework against alternative models, and conducted in-vivo experiments to compare its performance against established MWI approaches.
- We demonstrated highly efficient, high-resolution multi-parametric mapping at both 1 mm and 0.7 mm isotropic resolutions within 5 min and 10 min, respectively, utilizing multi-contrast and multi-slice zero-shot self-supervised learning (MZS-SSL) method for joint reconstruction of multi-contrast/-slice complex-valued images.
- We validated the translational potential of the framework through the assessment of multiple sclerosis lesions and investigated its ability to estimate cortical myelination.
- To promote reproducibility, the source code, including Pulseq (Layton *et al.*, 2017; Mancini *et al.*, 2020; Karakuzu *et al.*, 2022) sequence files, the reconstruction pipeline, and parameter estimation algorithms, is publicly available on GitHub: https://github.com/yutingchen11/MWF-MIMOSA.

# 2. METHODS AND MATERIALS

## 2.1 MWF-MIMOSA pulse sequence

The original MIMOSA sequence consists of a $T_2$ preparation module followed by one FLASH readout, and an inversion recovery (IR) preparation pulse followed by two FLASH readouts, and a monopolar MGRE module. The original sequence was primarily designed for single-compartment quantitative mapping. Consequently, the framework has limited sensitivity to the myelin water component, which is characterized by short relaxation times. Building upon this framework, the extended MWF-MIMOSA sequence was developed using MATLAB (MathWorks, Inc., Natick, MA, USA) on the open-source Pulseq platform (Layton *et al.*, 2017; Mancini *et al.*, 2020; Karakuzu *et al.*, 2022). To explicitly improve sensitivity to the myelin water compartment, an additional short echo time (TE = 20 ms) $T_2$-preparation module, followed by a single FLASH readout, was incorporated at the beginning of the sequence (Figure 1 (a)). This addition was motivated by previous studies showing that short-TE $T_2$-preparation improves MWI accuracy (Fujita *et al.*, 2025). Furthermore, a bipolar MGRE module with high readout efficiency and reduced echo spacing was implemented to replace the monopolar MGRE readout, thereby further enhancing myelin sensitivity and improving microstructure modeling (Shin *et al.*, 2019; Chan and Marques, 2020).

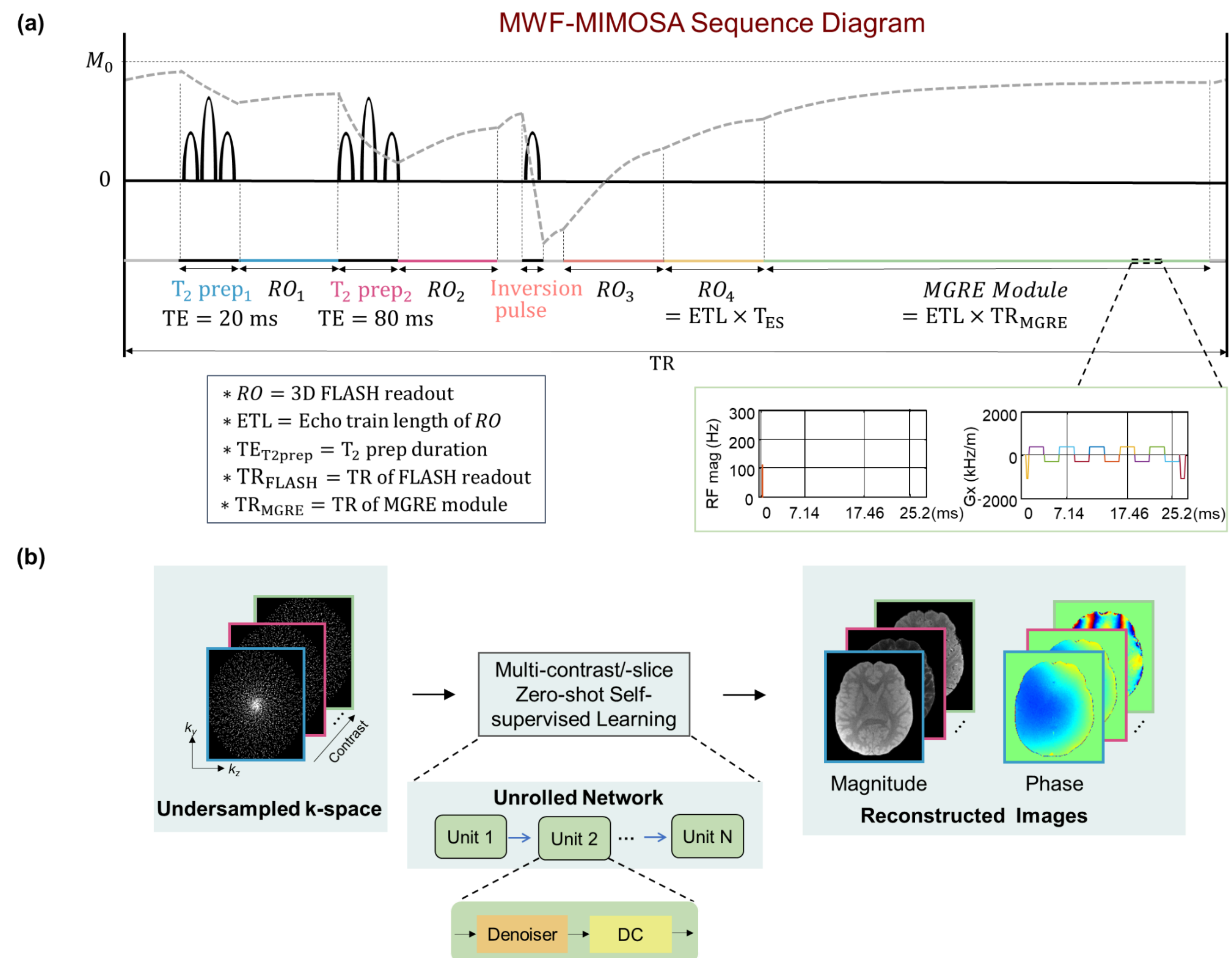

**Figure 1.** Overview of the MWF-MIMOSA framework. (a) Sequence diagram illustrating the acquisition modules. The sequence consists of a short TE (20 ms) $T_2$ preparation module ($T_2$ prep$_1$) followed by a FLASH readout ($RO_1$), a long TE (80 ms) $T_2$ preparation module ($T_2$ prep$_2$) followed by a FLASH readout ($RO_2$), an inversion pulse followed by two FLASH readouts ($RO_3$, $RO_4$), and a 3D multi-echo bipolar gradient echo (MGRE) module to generate $T_2^*$ and magnetic susceptibility contrasts. The dashed box details the MGRE module parameters for experiment at 1 mm isotropic resolution: TEs = [1.98, 4.56, 7.14, 9.72, 12.30, 14.88, 17.46, 20.04, 22.62, 25.20] ms and TR = 27.7 ms. Total sequence TR of MIMOSA is 6,956.5 ms. (b) Reconstruction pipeline. Undersampled multi-contrast k-space data are reconstructed using multi-contrast/-slice zero-shot self-supervised learning (MZS-SSL). The reconstruction network was implemented as an unrolled architecture with multiple units. Each unit consisted of a residual learning-based denoiser followed by a data-consistency layer (DC) solved using a conjugate gradient algorithm.

## 2.2 Self-supervised learning-based multi-contrast image reconstruction

Conventional reconstruction methods (Sodickson and Manning, 1997; Pruessmann *et al.*, 1999; Griswold *et al.*, 2002; Lustig, Donoho and Pauly, 2007) typically reconstruct each contrast separately and therefore do not fully exploit the complementary information available across contrasts, especially in high frequency k-space regions. Deep learning-based reconstruction methods have recently emerged as a powerful alternative for accelerated MRI (Knoll *et al.*, 2020; Heckel *et al.*, 2024; Safari *et al.*, 2026). By leveraging shared representations and image priors from the acquired data, they can better suppress undersampling artifacts and improve reconstruction quality. Among them, self-supervised learning (Yaman *et al.*, 2022; Jun *et al.*, 2024; Feng *et al.*, 2025, 2026; Li *et al.*, 2025) is particularly attractive because it eliminates the need for fully sampled reference data. In zero-shot self-supervised learning (ZS-SSL) (Yaman *et al.*, 2022), the acquired k-space data are divided into three disjoint subsets for training, validation, and loss computation, thereby obviating the need for reference data. MZS-SSL (Jun *et al.*, 2025; Chen *et al.*, 2026) extends this strategy by exploiting complementary information across contrasts and slices, enabling robust quantitative mapping under highly undersampled acquisitions.

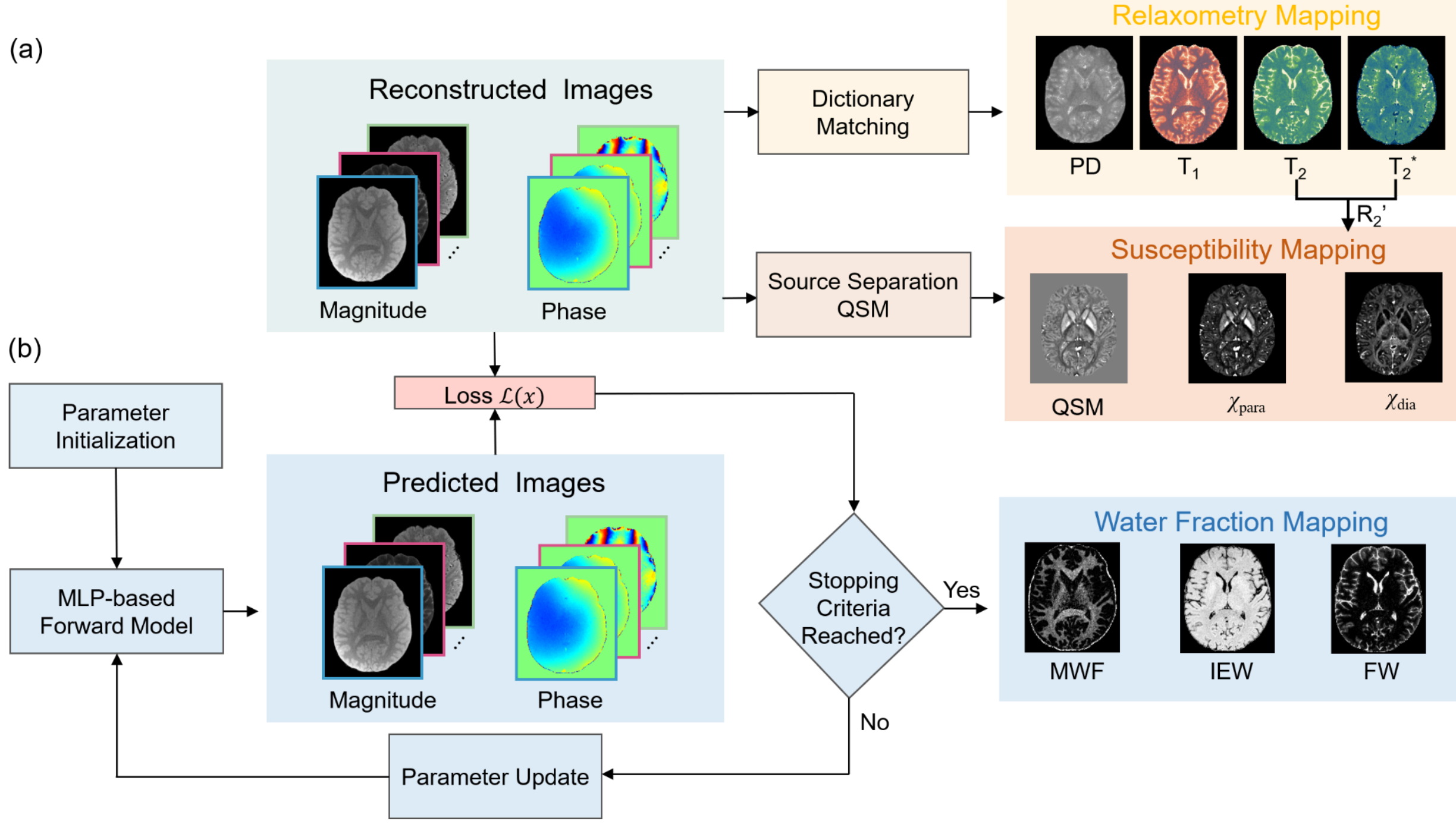

**Figure 2.** Quantitative parameter estimation pipeline of MWF-MIMOSA. (a) Single-compartment parameter estimation. Reconstructed magnitude images undergo dictionary matching to estimate $T_1$, $T_2$, and $T_2^*$ maps. Phase images are used further used for quantitative susceptibility mapping (QSM) and susceptibility source separation, incorporating an $R_2'$ map derived from $T_2$ and $T_2^*$. (b) GACELLE-based Myelin water fraction (MWF) estimation. MWF is estimated using a GACELLE-based multilayer perceptron (MLP) forward model, which iteratively updates parameters $x$ until convergence to yield the final water fraction maps.

In the proposed MWF-MIMOSA framework, MZS-SSL was used to jointly reconstruct whole-brain complex-valued images for all contrasts. Complementary acquisitions with different sampling patterns (Figure S1) were applied to readouts $RO_{1\text{–}4}$. In the MGRE module, the same sampling pattern was used across echoes to preserve the minimum echo spacing. The acquired undersampled k-space data was divided into three disjoint subsets using a random uniform distribution for training, validation, and loss calculation, respectively. The reconstruction network followed an unrolled design with multiple cascaded units, in which image prior learning and data fidelity were enforced in an alternating manner, as shown in Figure 1 (b). Specifically, each unit consisted of a residual learning-based denoiser followed by a data-consistency layer implemented with a conjugate gradient algorithm. The network was trained on an NVIDIA A100 GPU (NVIDIA, Inc., Santa Clara, CA, USA). The reconstructed multi-contrast complex-valued images were used for the quantitative parameter estimation as detailed in the next section.

## 2.3 Single compartment parameter mapping with MWF-MIMOSA

To account for the additional $T_2$-preparation module, the MWF-MIMOSA signal model was updated from the original MIMOSA model for the Bloch simulations. Other sequence parameters, including the TEs used for the MGRE module, were adjusted accordingly. A dictionary, generated via Bloch simulations using the updated MWF-MIMOSA signal model (Chen *et al.*, 2026) was subsequently used to estimate proton density (PD), $T_1$, $T_2$, and $T_2^*$ maps from the reconstructed magnitude images (Figure 2 (a)). The reconstructed phase images were first corrected for bipolar phase offsets (Liu *et al.*, 2011; Chan and Marques, 2021). After phase unwrapping (Robinson *et al.*, 2017; Dymerska *et al.*, 2021) and background-field removal (Wu, Li, Guidon, *et al.*, 2012; Zhou *et al.*, 2014), a tissue field map can be estimated (Wu, Li, Avram, *et al.*, 2012). This map was then used to estimate total QSM and separate the paramagnetic and diamagnetic susceptibility sources using χ-sepnet (Shin *et al.*, 2021; Kim *et al.*, 2025), incorporating an $R_2'$ map inferred from the estimated $T_2$ and $T_2^*$ maps ($R_2'=1/T_2^*-1/T_2$) (Figure 2 (a)). Having obtained these parameter maps, the next section details the partial volume modeling required for MWF estimation.

## 2.4 MWF mapping with MWF-MIMOSA

### 2.4.1 Three-compartment signal model

For MWF estimation, a three-compartment model comprising myelin water (MW), intra- and extra-axonal water (IEW), and free water (FW) was incorporated into the MWF-MIMOSA signal model.
Let $C$ = {MW, IEW, FW} denote the set of modeled water compartments, and $x$ represent the comprehensive set of unknown model parameters to be estimated:

$$x = \left[S_0, \Delta\omega_b, \phi, \left\{f_c, T_{1,c}, T_{2,c}, T_{2,c}^{*}, \Delta\omega_c\right\}_{c\in C}\right], \quad (1)$$

where $S_0$ denotes the PD scaling factor, which can be estimated by dictionary matching as described in the previous section. $\Delta\omega_b$ is the background frequency offset originating from the macroscopic field inhomogeneity, and $\phi$ is the initial phase at TE = 0 ms coming from the receive/transmit phase offset. The initial values of $\Delta\omega_b$ and $\phi$ parameters were obtained from the images acquired in the MGRE module using the MEDI toolbox (Liu *et al.*, 2011). The variables $f_c$, $T_{1,c}$, $T_{2,c}$, $T^*_{2,c}$, and $\Delta\omega_c$ represent the compartment fraction, relaxation parameters, and frequency shift for compartment $c$, respectively.

The signals predicted using the MWF-MIMOSA forward model can be described as the sum of compartment-wise signals based on Bloch simulations as follows:

$$\hat{S}(t;x) = S_0 \left( \sum_{c \in C} f_c \mathcal{B} \left( t; T_{1,c}, T_{2,c}, T_{2,c}^*, \Delta\omega_c \right) \right) e^{i\Delta\omega_b t} e^{i\phi} \tag{2}$$

where $\mathcal{B}$ represents the Bloch simulation operator based on the MWF-MIMOSA sequence timing, and *t* stands for the echo time.

We assume the free water compartment is a homogenous medium such that $\Delta\omega_{\mathrm{FW}}$ = 0, and we further fix the relaxometric parameters of the free water ($T_{1,FW}$ = 4,500 ms, $T_{2,FW}$ = 500 ms, $T^*_{2,FW}$ = 500 ms) (Chen *et al.*, 2019; Deshmane *et al.*, 2019; H. G. Kim *et al.*, 2023). Therefore, MWF can be estimated by solving the following optimization problem:

$$\begin{gathered} \mathcal{L}(x) = \|S - \hat{S}(x)\|_1 + \mathcal{R}(x) \\ \hat{x} = \arg\min_x \mathcal{L}(x) \\ \text{s.t. } \sum_{c \in C} f_c = 1, \quad f_c \geq 0, \ \forall c \in C \end{gathered} \tag{3}$$

where *S* is the measured complex signal vector, and *R*(x) denotes the total variation (TV) regularization incorporated for $f_{MW}$ only to improve the spatial smoothness and stability of the MWF solution. The regularization parameter $\lambda$ was selected based on pilot experiments to balance noise suppression with the preservation of spatial variations in the MWF maps, and a fixed value of $3 \times 10^{-4}$ was used for all datasets. The $L_1$-norm loss was utilized in the data fidelity term, as it has been shown to provide more stable convergence than the conventional $L_2$-norm loss (Chan *et al.*, 2025). The optimization is subject to physical constraints, ensuring that the compartment fractions are non-negative and sum to unity.

### 2.4.2 GPU acceleration

Existing MGRE-based MWI methods estimate MWF and relaxometric parameters jointly using voxel-wise least-squares optimization. However, this fitting is often computationally expensive and sensitive to initialization. MWF-MIMOSA requires Bloch simulations to generate signal evolutions in Eq. (2), leading to a computationally intensive optimization problem. As there is no closed form expression for the MWF-MIMOSA signal model (Eq. (2)), computationally intensive Bloch simulations must be performed multiple times at each iteration. GACELLE (Chan *et al.*, 2025), an open source GPU-accelerated toolbox, provides a practical framework for accelerated model-based parameter estimation. It supports multiple optimisers, configurable loss functions, spatial regularisation, while requiring only the user-defined forward signal model. These features make it well suited for multi-compartment MWF estimation. Therefore, GACELLE was used to solve Eq. (3) by performing parallel computation across the voxels in each slice. To enable efficient parallel computing on GPU with GACELLE, a physics-informed MLP neural network was trained as a surrogate for the Bloch simulation, reducing the computational cost by more than two orders of magnitude.

With this framework, the MWF estimation time for each slice was reduced from several hours to approximately 10 seconds for a typical slice with around 15,000 voxels inside the brain mask (≥ 100-fold acceleration). Further details regarding the MLP architecture, training procedure, validation, and ablation studies are provided in Supplementary Materials.

As illustrated in Figure 2(b), the MLP-based forward model takes the current parameter estimates as input and predicts the corresponding complex signals, from which the optimization loss in Eq. (3) was computed. The parameters were then iteratively updated within GACELLE until the predefined stopping criteria were reached, when either the loss or its gradient fell below the empirically determined threshold of $10^{-6}$.

## 2.5 Simulations

In the simulations, typical three-compartmental $T_1$, $T_2$, and $T_2^*$ values at 3T were assigned to the BrainWeb (Collins *et al.*, 1998; Kwan, Evans and Pike, 1999) digital brain phantom. The assigned values were: $T_{1,MW/IEW/FW}$ = 150/1000/4500 ms, $T_{2,MW/IEW/FW}$ = 15/70/500 ms, $T^*_{2,MW/IEW/FW}$ = 10/50/500 ms.

The first simulation was designed to investigate how various models performed at different MWF regimes/tissue types using an in silico head phantom comprising white matter (WM), gray matter (GM), and cerebrospinal fluid (CSF). To evaluate the impact of three modeling strategies for MWF estimation, namely fixed compartment-specific relaxation parameters, conventional MGRE-only estimation, and the proposed joint-estimation framework, we compared the following three models:

- $Model_{Prep}$, which uses images acquired after the preparation modules in MWF-MIMOSA containing only $T_1$- and $T_2$-weighted contrasts ($RO_{1–4}$ and the first echo from the MGRE module). For MWF estimation, the ground truth compartmental $T_{1,MW/IEW/FW}$ and $T_{2,MW/IEW/FW}$ values were used as known and fixed during fitting.
- $Model_{MGRE}$, which uses only the images from the MGRE module. Analogous to conventional MGRE-based MWF techniques, a multi-compartment exponential model incorporating $T_2^*$ decay and frequency evolution was fitted using nonlinear least squares optimization (Nam *et al.*, 2015; Chan and Marques, 2020).
- $Model_{MWF-MIMOSA}$, which uses all acquired images and jointly estimates MWF together with compartmental relaxation parameters.

The compartmental fractions in the phantom were: MWF = 0.10, IEW = 0.88, and FW = 0.02 in WM, MWF = 0.02, IEW = 0.96, and FW = 0.02 in GM, and MWF = 0, IEW = 0, and FW = 1 in CSF. The hollow cylinder fiber model (HCM) (Wharton and Bowtell, 2012; Sati *et al.*, 2013) was used to simulate orientation-dependent compartmental frequency shifts, assuming a WM fiber orientation of 45°.

To investigate if our proposed method improves the estimation robustness, we varied the initialization condition in our second simulation. We randomly generated 10 different initializations for all models such that accuracy and precision can be assessed by uniformly sampling the MWF start point from 0.05 to 0.15. Accuracy was assessed using the mean absolute error (MAE) of the estimated MWF relative to the ground truth. Precision was quantified by the mean standard deviation (MSTD) of the estimated MWF across the 10 initializations for each model. To further evaluate the robustness of MWF-MIMOSA across fiber orientations, two extreme fiber orientations (0° and 90°) were also simulated across a range of MWF = [0.05:0.05:0.2] in WM.

The sequence parameters used in Bloch simulations were: echo train length (ETL) = 127, echo spacing (ESP) = 5.8 ms, TE = 2.29 ms, flip angle (FA) = 4˚, TR of MGRE = 27.7 ms, TEs of MGRE = [1.98, 4.56, 7.14, 9.72, 12.30, 14.88, 17.46, 20.04, 22.62, 25.20] ms, and whole block TR = 6,956.5 ms. Gaussian noise was added to achieve an SNR of 40 dB.

## 2.6 In-vivo validation and statistical analysis

### 2.6.1 Hardware and institutional approvals

All in-vivo experiments were performed with the approval of the Institutional Review Board on 3T scanners (MAGNETOM Prisma and Cima.X, Siemens Healthineers, Forchheim, Germany) equipped with a 64-channel head receiver coil. SVD coil compression (Zhang *et al.*, 2013) was used to compress the data to 20-channels prior to reconstruction to reduce reconstruction time and computation demands.

### 2.6.2 Acquisition protocols

**MWF-MIMOSA at 1 mm isotropic resolution:** To compare the MWF maps from different models, one subject (female; age, 26 years) was scanned using the MWF-MIMOSA sequence with the following imaging parameters: field of view (FOV) = 240×224×192 $mm^3$, resolution = 1 mm isotropic, ETL = 127, ESP = 5.8 ms, TR of MGRE = 27.7 ms, TEs of MGRE = [1.98, 4.56, 7.14, 9.72, 12.30, 14.88, 17.46, 20.04, 22.62, 25.20] ms, FA = 4˚, total sequence TR = 6,055 ms, acceleration factor (R) = 6, total acquisition time (TA) = 6 min 4 sec.

For subsequent region of interest (ROI) analysis, five additional subjects (two male and three female; mean age, 28 years) were scanned using identical protocols, with the exception of higher acceleration (R = 8) and a corresponding TA of 4 min 55 sec. One patient with relapsing-remitting multiple sclerosis (MS) (female; age, 54 years; disease duration approximately 10 years) was scanned using the same 4 min 55 sec protocol (Cima.X, Siemens Healthineers, Forchheim, Germany).

**MWF-MIMOSA at 0.7 mm isotropic resolution**: To demonstrate high-resolution mapping capabilities, a 0.7 mm isotropic resolution scan was performed on one of the five subjects (male; age, 29) with the following protocol: FOV = 200×224×192 $mm^3$, resolution = 0.7 mm isotropic, ETL = 119, ESP = 5.8 ms, TR of MGRE = 29.4 ms, TEs of MGRE = [2.38, 5.38, 8.38, 11.38, 14.38, 17.38, 20.38, 23.38, 26.38] ms, FA = 4˚, total sequence TR sequence = 6,055 ms, R = 8, TA = 9 min 48 sec.

$B_1^+$ maps were acquired using a product turbo-FLASH sequence (Chung *et al.*, 2010) to correct for transmit field inhomogeneity in both the single-compartment and three-compartment signal models.

### 2.6.3 Comparison methods

We assessed if the MWF maps derived from MWF-MIMOSA were comparable to the established methods, including MCR-MWI (Chan and Marques, 2020), a variable-flip-angle MGRE method incorporating magnetization-exchange modeling, and GRASE (Piredda, Hilbert, Canales-Rodríguez, *et al.*, 2021), a multi-echo gradient-and-spin-echo method.

**MCR-MWI:** FOV = 240×240×192 mm$^3$, resolution = 1.5 mm isotropic, FA = [10˚, 20˚, 50˚], R = 5 with CAIPI acceleration (z-shift = 2), and TA = 10 min 27 sec.

**GRASE:** FOV = 240×210×144 mm$^3$, resolution = 1x1x4 mm$^3$; 32 echoes with a minimum TE and echo spacing of 13.12 ms, TR = 1,250 ms, FA = 185˚, R = 6 with CAIPI acceleration, and TA = 9 min 15 sec.

### 2.6.4 Image processing and statistical analysis

For quantitative ROI analysis, all MWF maps were nonlinearly registered to the 1 mm MNI152 space (Grabner *et al.*, 2006; Fonov *et al.*, 2011) using ANTs (Avants *et al.*, 2008, 2011; Klein *et al.*, 2009; Murphy *et al.*, 2011). To quantitatively compare the MWF maps of MWF-MIMOSA with MCR and GRASE, ROI analysis was performed in five major WM ROIs: Forceps major (FMj), Forceps minor (FMn), Corticospinal tract (CST), Cingulum (CG), and Sagittal stratum (SS)) using the JHU ICBM-DTI-81 WM atlas (Mori *et al.*, 2008).

Pearson's correlation coefficients (*r*) and p-values were computed to evaluate the correlations between the MWF values obtained with MWF-MIMOSA and comparison methods across 17 major WM ROIs. These ROIs included the Genu, Body, Splenium of the Corpus Callosum, as well as the Corticospinal Tract (R/L), Anterior and Posterior Limbs of the Internal Capsule (R/L), Inferior and Superior Longitudinal Fasciculi (R/L), Cingulum (R/L), and Uncinate Fasciculus (R/L).

Finally, as an exploratory analysis, we moved beyond conventional WM analysis to study the MWF in the cortical gray matter. MWF maps were mapped to the cortical surface for each subject using Freesurfer (Fischl, 2012). To enable cross-subject comparison, the surface maps were then registered to a common template surface (fsaverage; Fischl, 2012), and the effects of cortical curvature were regressed out. Group-average MWF was subsequently computed on the group-average midthickness surface.

# 3. RESULTS

## 3.1 Simulation evaluation

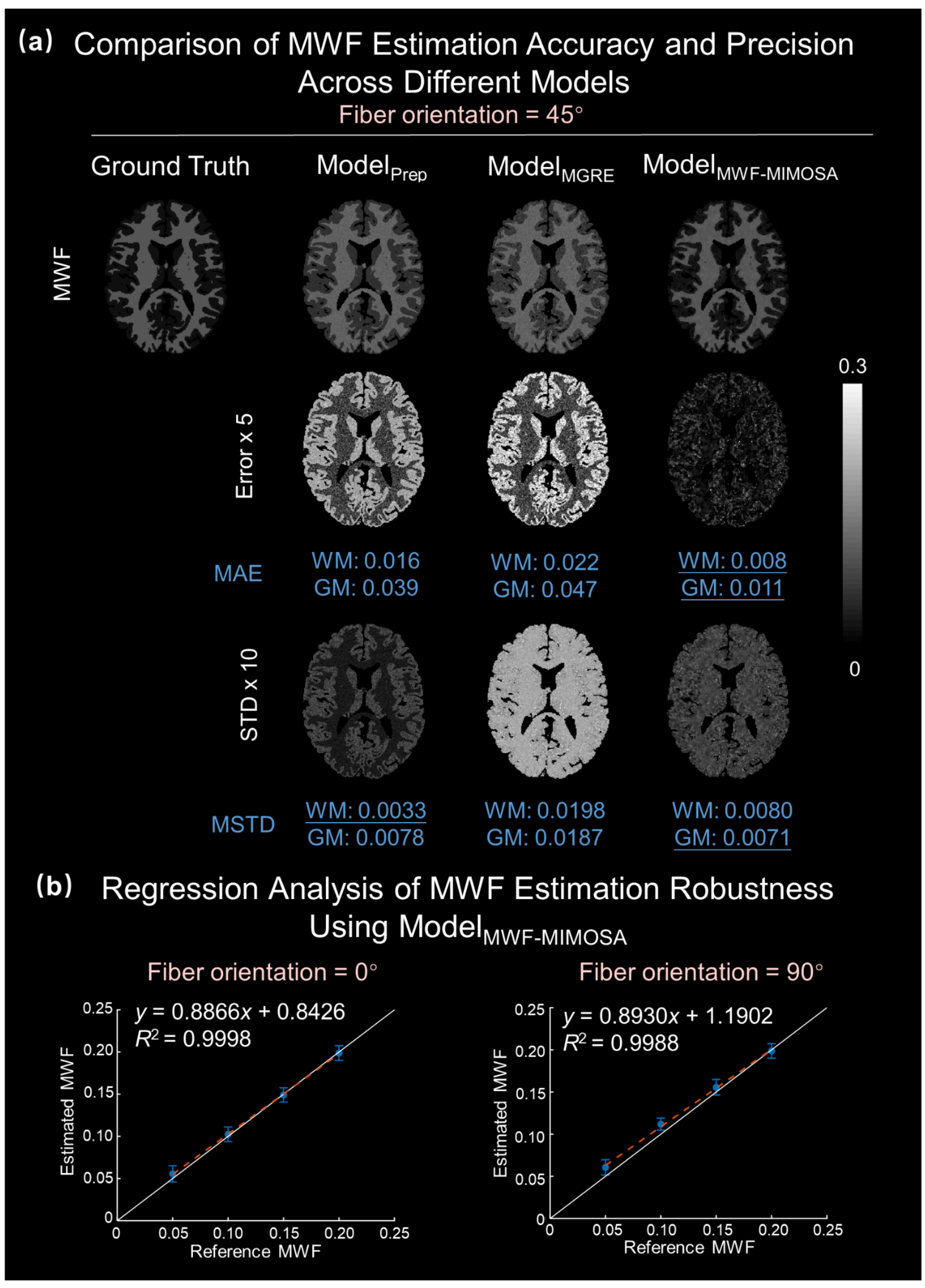

**Figure 3.** Numerical simulation results. (a) Comparison of mean myelin water fraction (MWF) maps across 10 initializations using Model$_{Prep}$, Model$_{MGRE}$, and the proposed Model$_{MWF-MIMOSA}$ at a 45˚ fiber orientation. The corresponding error and standard deviation maps are displayed with 5x and 10x scaling, respectively. The lowest mean absolute error (MAE) and mean standard deviation (MSTD) for each tissue type are underlined. (b) Regression analysis of MWF estimation robustness using Model$_{MWF-MIMOSA}$, evaluated across MWF values ranging from 0.05 to 0.2 at parallel (0˚) and perpendicular (90˚) fiber orientations. The error bar shows the MSTD within WM.

Figure 3 (a) presents the mean MWF maps alongside the corresponding error and MSTD maps across initializations, for the evaluated models. $Model_{Prep}$ showed the lowest initialization sensitivity in WM (MSTD = 0.0033), likely because it used fixed relaxometric parameters that matched to the ground truth in the simulation settings. However, it systematically overestimated the MWF and showed bias in both WM and GM, yielding MAE of 0.016 and 0.039, respectively. This bias is consistent with the model's inability to account for $T_2^*$ decay and frequency offsets (Nam *et al.*, 2015). In addition, the MSTD of $Model_{Prep}$ increased when the fixed relaxometric parameters were varied across initializations (Figure S4), indicating high sensitivity to parameter mismatch.

Conversely, $Model_{MGRE}$ was substantially more sensitive to initialization, producing higher MSTD values in both WM (0.0198) and GM (0.0187). It also resulted in large estimation errors, particularly in GM (MAE = 0.022; WM MAE: 0.047), reflecting the ill-posed nature of simultaneously fitting compartment fractions, $T_2^*$, and frequency parameters from MGRE data alone.

In contrast, the proposed MWF-MIMOSA model ($Model_{MWF\text{-}MIMOSA}$) achieved the lowest MAE in both WM (0.008) and GM (0.011), along with the lowest GM MSTD (0.0071). Although its WM MSTD (0.0080) was higher than that of $Model_{Prep}$, it provided the best overall balance between accuracy and precision; therefore, it was used for all subsequent analyses. Figure 3 (b) demonstrated that MWF-MIMOSA maintained strong agreement with the ground-truth MWF under extreme parallel and perpendicular fiber orientations, yielding slopes of 0.8866 ($R^2$ = 0.9998) at 0˚ and 0.8930 ($R^2$ = 0.9988) at 90˚.

## 3.2 In-vivo validation and statistical analysis

### 3.2.1 In-vivo model comparison

The in-vivo results supported the simulation findings (Figure 4). MWF estimated using $Model_{Prep}$ was consistently the highest among the three models, whereas $Model_{MGRE}$ produced low-SNR MWF maps due to the ill-posed nature of the estimation. In contrast, $Model_{MWF\text{-}MIMOSA}$ showed the lowest standard deviation among all models (MSTD, $Model_{MWF\text{-}MIMOSA}$ vs. $Model_{Prep}$ vs. $Model_{MGRE}$ = 0.007 vs. 0.012 vs. 0.013), indicating the highest robustness to initialization. Figure 5 displays the whole-brain $T_1$, $T_2$, $T_2^*$, QSM, paramagnetic susceptibility, diamagnetic susceptibility, and MWF maps acquired at 1 mm isotropic resolution in 5 min. Additional compartmental parameter maps were presented in Figure S5.

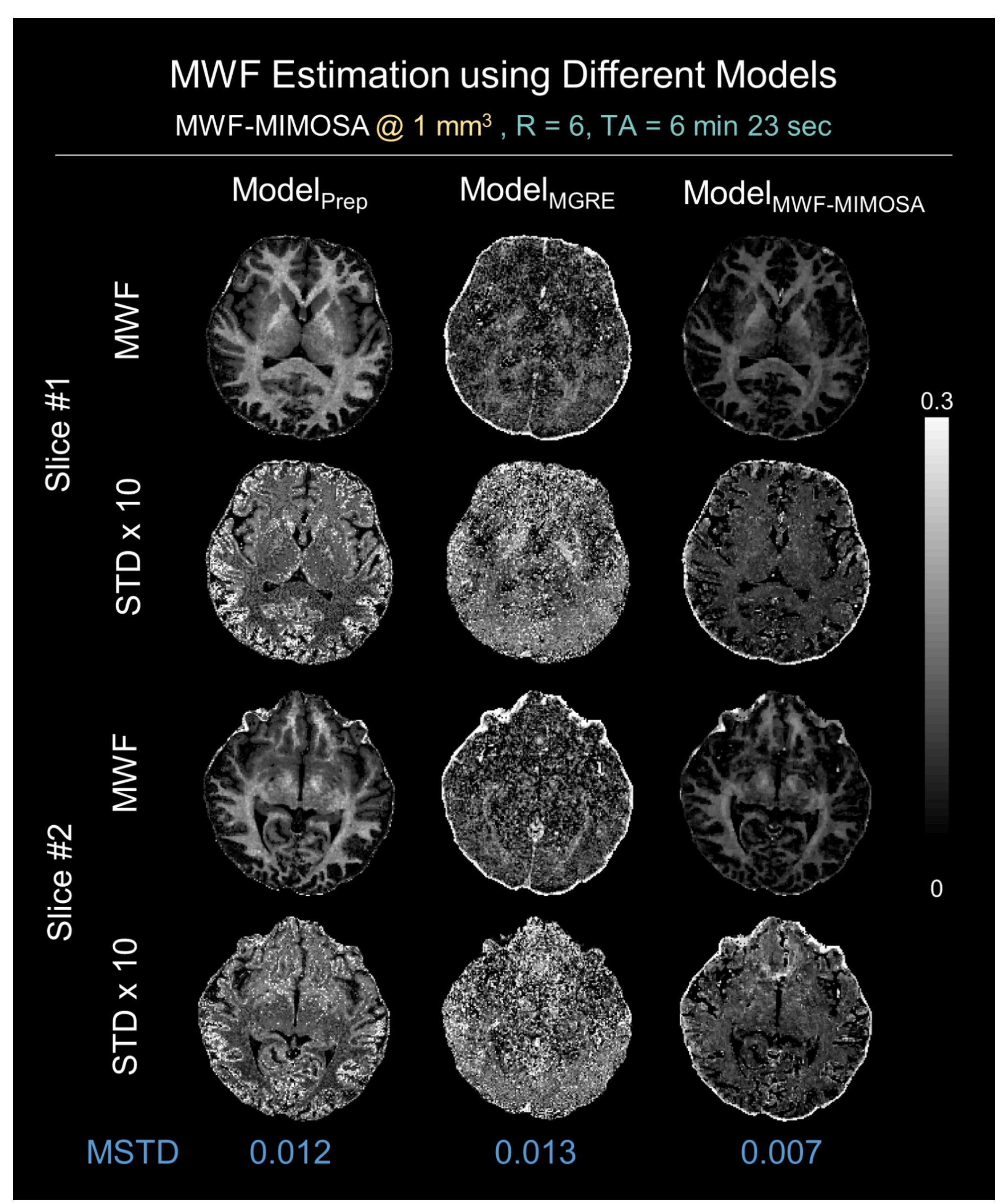

**Figure 4.** In-vivo evaluation of myelin water fraction (MWF) estimation models. Comparison of mean MWF maps across 10 initializations for $Model_{Prep}$, $Model_{MGRE}$, and $Model_{MWF\text{-}MIMOSA}$. Rows 1 and 3 present two representative slices from a single subject. Rows 2 and 4 show the corresponding standard deviation maps across 10 initializations, scaled by a factor of 10 for visibility. For each model, the mean standard deviation (MSTD) is reported.

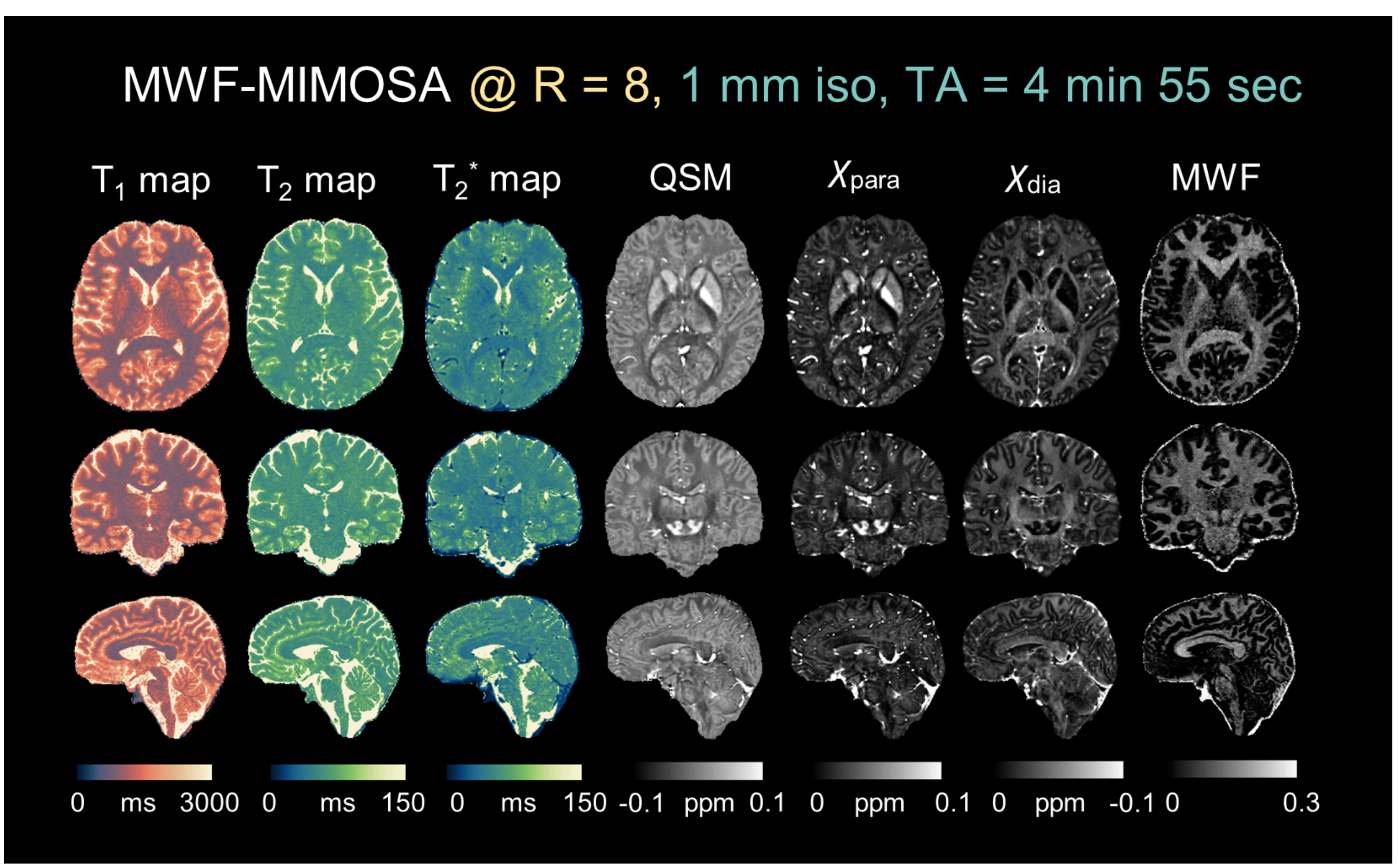

**Figure 5.** In-vivo multiparametric mapping. Simultaneous whole-brain estimation of $T_1$, $T_2$, $T_2^*$, quantitative susceptibility mapping (QSM), susceptibility source separation ($X_{para}$ and $X_{dia}$) and myelin water fraction (MWF) at 1 mm isotropic resolution, acquired in 4 min 55 sec using MWF-MIMOSA.

### 3.2.2 Comparison of MWF estimation methods

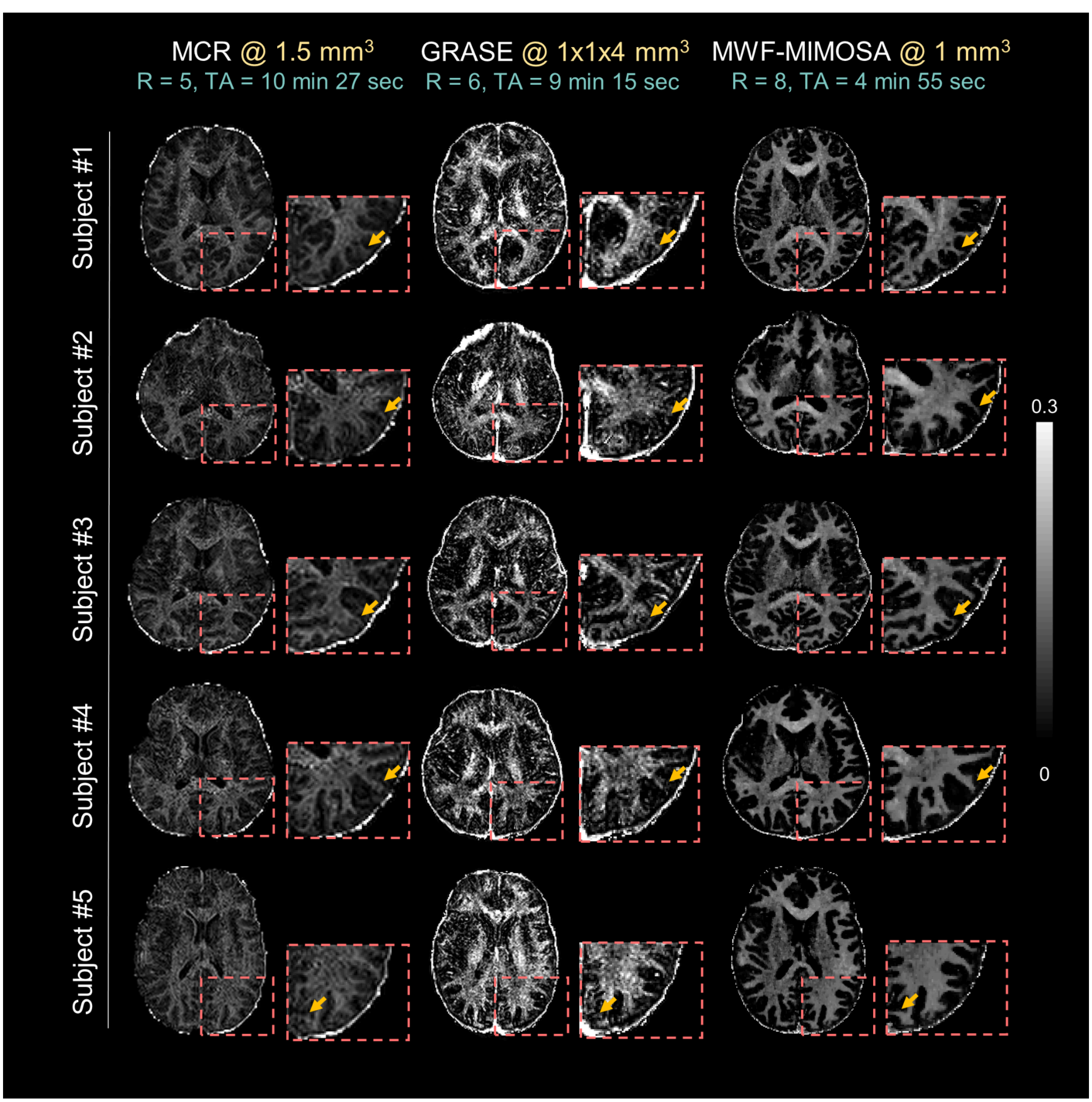

**Figure 6.** Qualitative comparison of in-vivo myelin water fraction (MWF) mapping. MWF maps from five subjects obtained using MCR, GRASE, and MWF-MIMOSA. MWF-MIMOSA achieves higher spatial resolution in a shorter acquisition time. Orange arrows highlight the improved delineation of white matter fiber bundles, particularly at the distal ends of major fiber tracts, in the MWF-MIMOSA maps.

Figure 6 compares the MWF maps of five subjects obtained with MCR, GRASE, and the proposed MWF-MIMOSA framework. MWF-MIMOSA achieved improved image quality and higher spatial resolution in a shorter acquisition time compared to MCR and GRASE. Notably, it can be appreciated that WM fiber bundles were more clearly delineated in the MWF-MIMOSA maps due to the improvement in resolution, particularly at the distal ends of major fiber tracts (orange arrows in Figure 6).

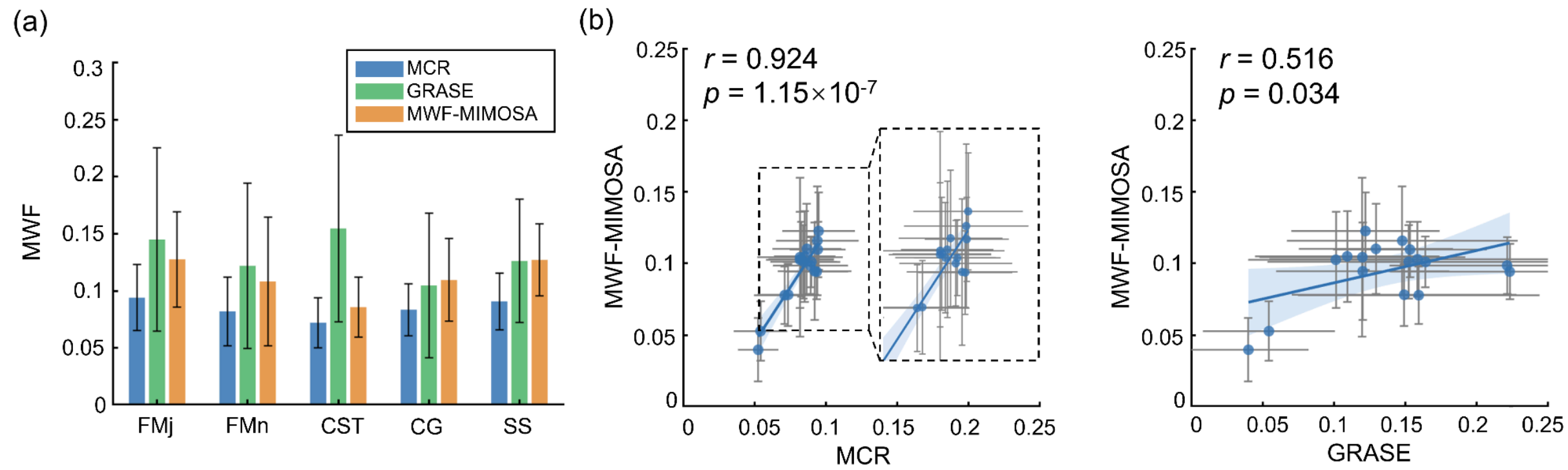

**Figure 7.** Quantitative region of interest (ROI) analysis. **(a)** Mean and standard deviation of myelin water fraction (MWF) values across five subjects within five major white matter ROIs for MWF-MIMOSA, MCR, and GRASE. **(b)** Correlation analysis of MWF estimates across 17 white matter ROIs between MWF-MIMOSA and other methods. Each data point represents the mean MWF value from the five subjects for a specific ROI, with error bars indicating the standard deviation. Blue shading denotes stronger positive correlations.

In the ROI-based analysis (Figure 7(a)), MWF-MIMOSA demonstrated good agreement with the other methods. Across the five ROIs, all three methods showed a similar relative pattern, with higher MWF in FMj and SS and lower MWF in CST. GRASE tended to estimate higher MWF than the other methods in the CST, likely due to $T_2$* confounding effects. The individual subject ROI analyses are presented in Figure S6.

Correlation analysis across the five subjects and 17 major WM ROIs (Figure 7(b)) showed a strong, statistically significant correlation between MWF-MIMOSA and MCR ($r$ = 0.924, $p$ = $1.15\times10^{-7}$). A moderate-to-strong, statistically significant correlation was also observed between MWF-MIMOSA and GRASE ($r$ = 0.516, $p$ = 0.034). Furthermore, MWF-MIMOSA demonstrated reduced standard deviations across ROIs compared to GRASE (indicated by the error bars).

### 3.2.3 Cortical myelin analysis

Figure 8 visualizes the right-hemisphere cortical MWF maps for the five individual subjects alongside the group average. The spatial distribution of cortical MWF was broadly consistent across subjects, revealing elevated MWF in the primary and early unimodal areas, including the primary motor cortex (M1), primary somatosensory cortex (S1), middle temporal visual area (MT), and primary auditory cortex (A1), and reduced MWF in multimodal association regions. This distribution aligns with known cortical architecture (Glasser and Van Essen, 2011; Gordon, May and Nelson, 2019). We noted the reduced MWF dynamic range, from 1% to 3%, in this figure.

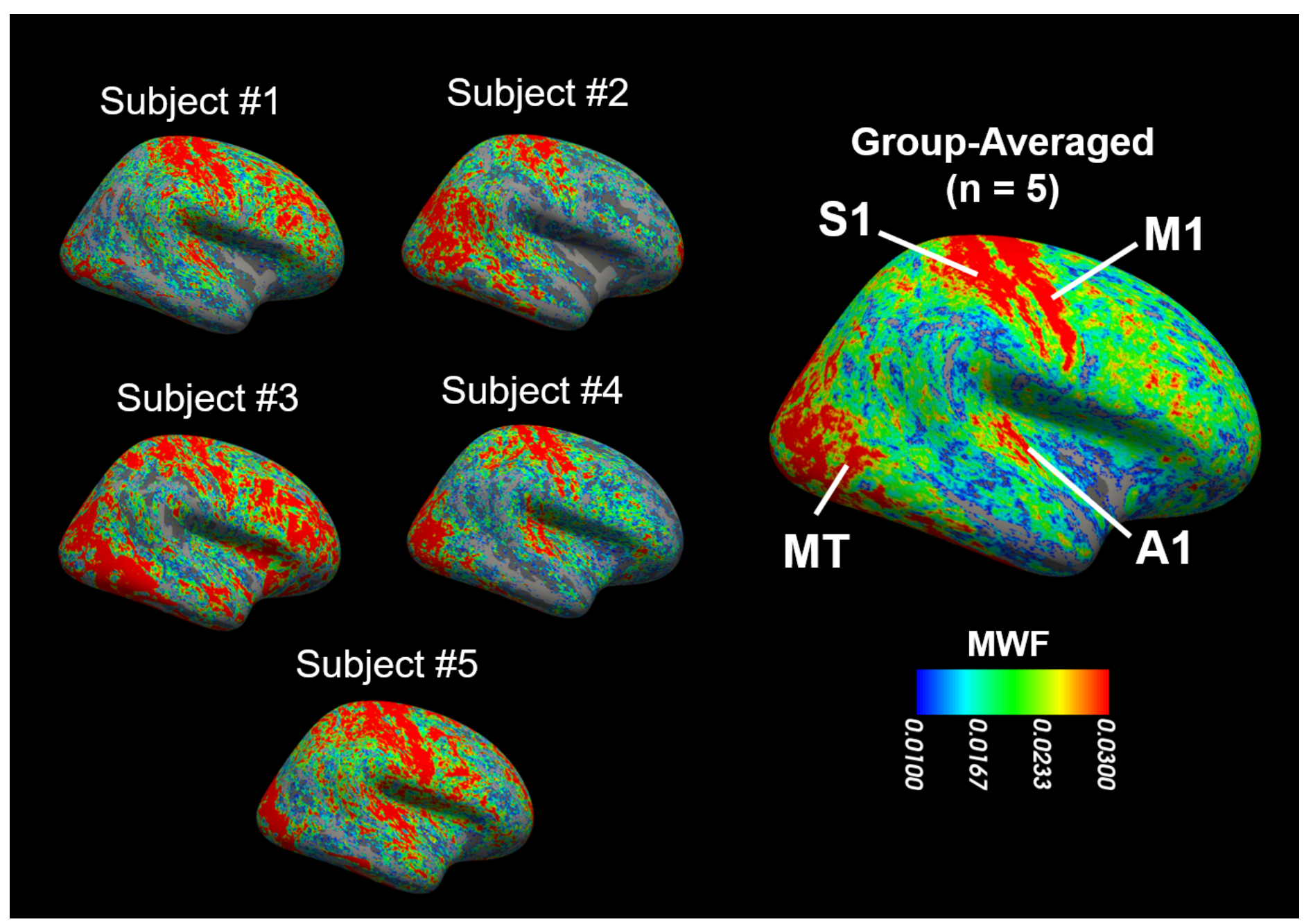

**Figure 8.** Cortical myelin mapping. Surface visualizations of right hemisphere cortical myelin water fraction (MWF) maps for five individual subjects and the resulting group average. Regions with relatively high myelination, including the primary motor cortex (M1), primary somatosensory cortex (S1), middle temporal visual area (MT), and primary auditory cortex (A1), are indicated.

### 3.2.4 High-resolution quantitative mapping and clinical application

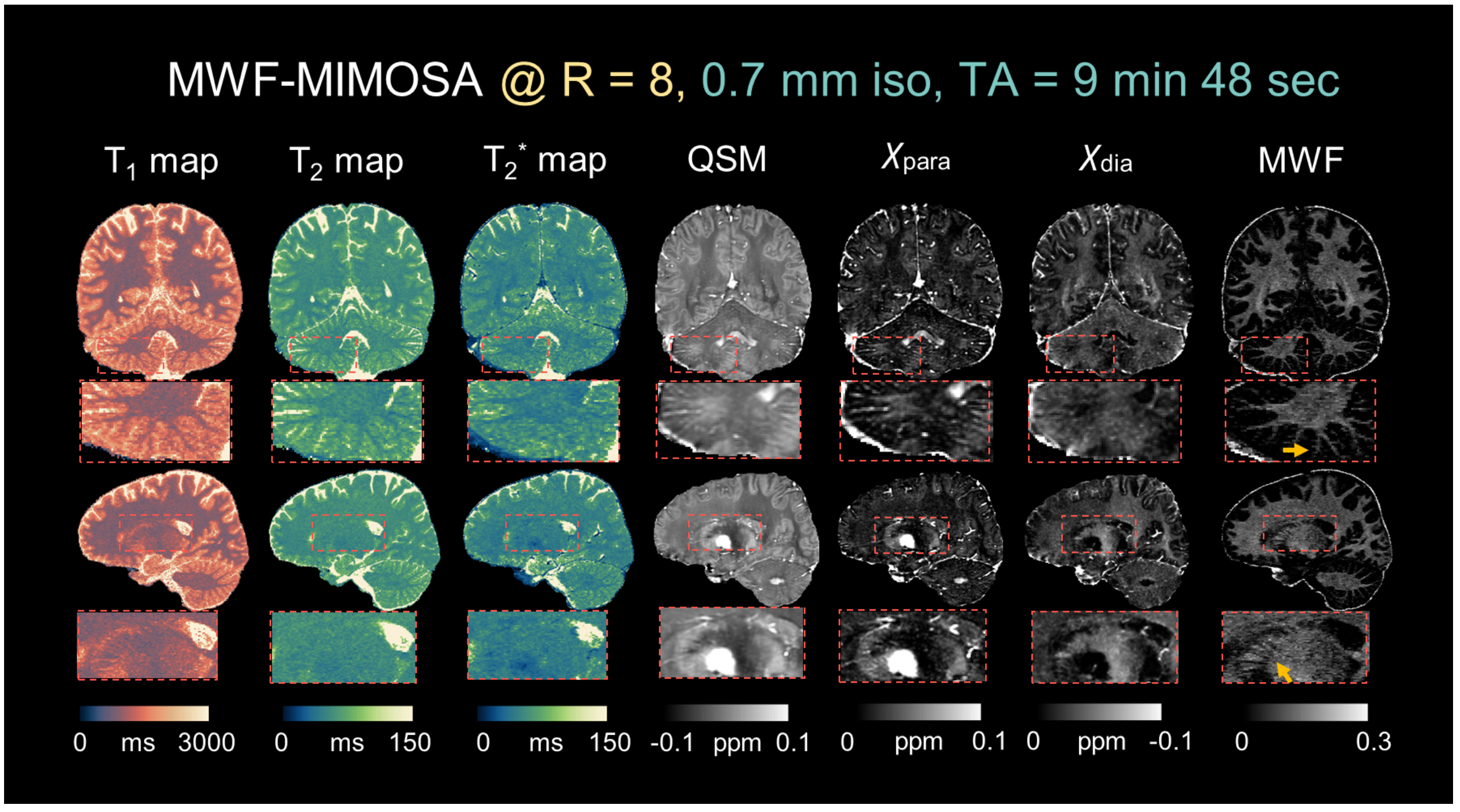

**Figure 9.** Whole-brain quantitative maps obtained using MWF-MIMOSA at 0.7 mm isotropic resolution in 9 min 48 sec. Yellow arrows indicate areas where the 0.7 mm resolution enables enhanced visualization of subtle structures, such as the internal capsule and cerebellar white matter fibers.

Figure 9 shows the multi-parametric maps acquired with MWF-MIMOSA at 0.7 mm isotropic resolution, enabling enhanced visualization of subtle brain structures. As indicated by the yellow arrows, the internal capsule and cerebellar WM fibers were more clearly resolved.

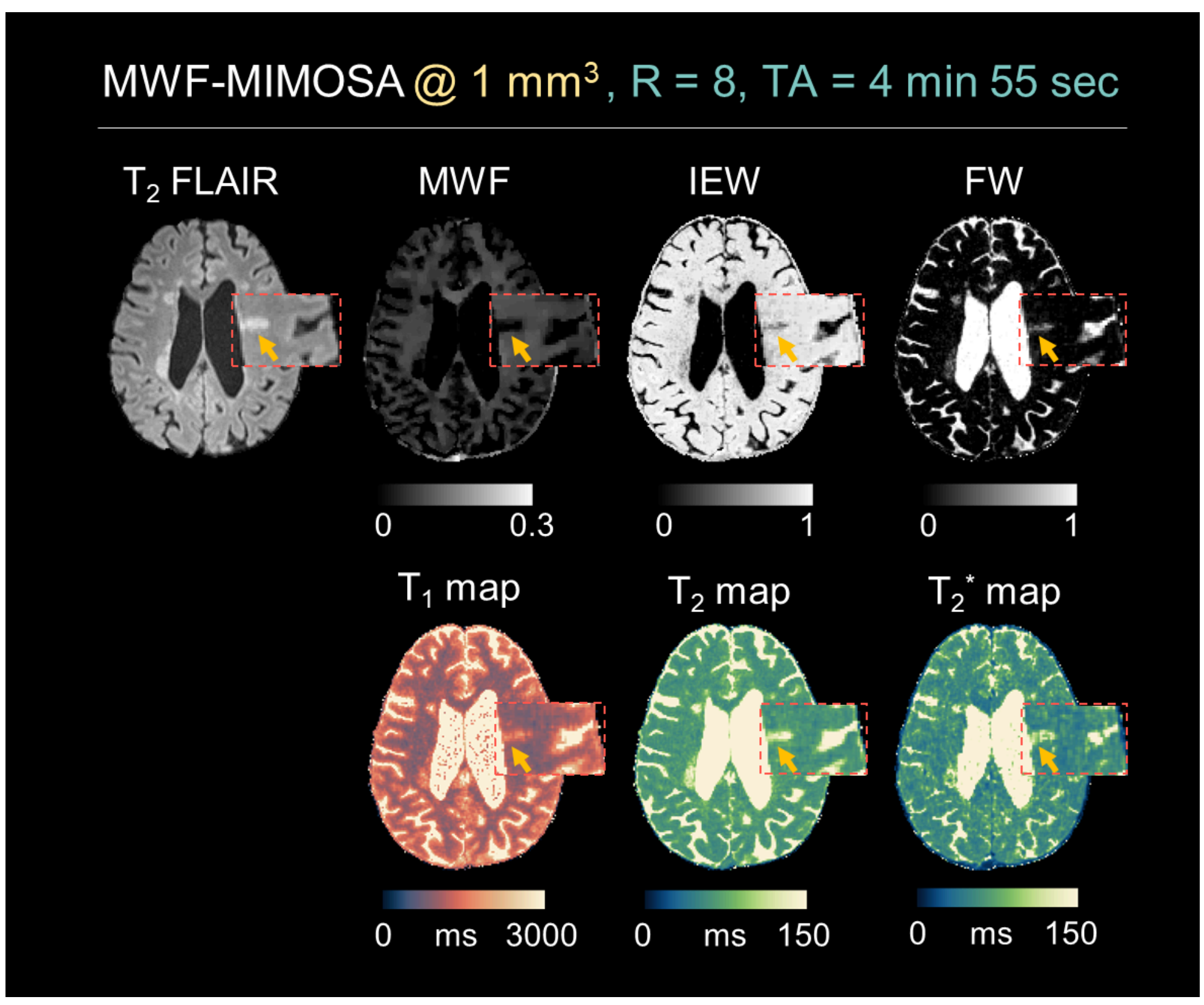

**Figure 10.** Clinical application in multiple sclerosis (MS). Clinical $T_2$-FLAIR images and corresponding quantitative maps obtained with MWF-MIMOSA 1 mm isotropic resolution protocol from a patient with MS. Yellow arrows highlight MS lesions, which exhibit reduced myelin water fraction (MWF) and increased intra- and extra-axonal water (IEW), and free water (FW), alongside with prolonged $T_1$, $T_2$, and $T_2^*$ relaxation times.

To demonstrate clinical feasibility, Figure 10 shows the quantitative maps obtained with 1 mm isotropic resolution MWF-MIMOSA alongside corresponding clinical $T_2$-FLAIR images from a patient with MS. The arrows highlight an MS lesion that appears hyperintense on $T_2$-FLAIR. The lesion was consistently identified on the MWF-MIMOSA and was characterized by reduced MWF, increased IEW and FW, and prolonged $T_1$, $T_2$, and $T_2^*$ relaxation times, suggesting predominant demyelination and increased tissue water content.

# 4. Discussion

In this work, we proposed MWF-MIMOSA, a novel framework for the simultaneous and efficient mapping of $T_1$, $T_2$, $T_2^*$, susceptibility source separation, and MWF. By employing MZS-SSL, multi-contrast and multi-slice zero-shot self-supervised learning, we achieved joint reconstruction of complex-valued images from acquisitions with complementary sampling without requiring an external training dataset. Furthermore, the integration of a MLP within the GACELLE framework bypassed computationally demanding Bloch simulations, yielding a >100-fold acceleration in MWF estimation. Simulations demonstrated that the proposed framework improves the accuracy and precision of MWF estimation compared to different models. In-vivo experiments further showed strong agreement with established MWI approaches, while simultaneously providing complementary quantitative maps at higher spatial resolutions and in shorter scan times. Specifically, we achieved multi-parametric mapping in 5 min at 1 mm isotropic resolution and in 10 min at 0.7 mm isotropic resolution, demonstrating the feasibility of high-resolution imaging for visualizing subtle brain structures. Moreover, cortical analyses identified heavily myelinated regions consistent with prior literature, supporting the potential of MWF-MIMOSA for in-vivo characterization of cortical myeloarchitecture. An initial application in a patient with MS showed that the lesions were further characterized by complementary information across different quantitative maps, highlighting the potential of MWF-MIMOSA for comprehensive tissue assessment.

MWF-MIMOSA aims to improve the estimation of myelin-related parameters by combining multiple myelin-sensitive contrasts into a single acquisition. Conventional GRE-based MWI primarily relies on differences in $T_2^*$ decay and frequency offsets to separate myelin water from other water pools. Because the corresponding multi-compartment fitting problem is ill-conditioned (Nam *et al.*, 2015), the estimation remains highly sensitive to MWF initialization. Incorporating variable-flip-angle GRE data can mitigate $T_1$-related confounds and improve SNR (Chan and Marques, 2020; Chan, Chamberland and Marques, 2023), though at the cost of acquiring multiple acquisitions that extend the total scan time. Alternatively, methods exploiting the short-$T_1$ or short-$T_2$ properties of myelin improve robustness either by reducing the fitted degrees of freedom (Chen *et al.*, 2019; Deshmane *et al.*, 2019) or by directly suppressing long-$T_1$ components (Oh *et al.*, 2013; Liao *et al.*, 2023). However, these simplifications can introduce bias when compartmental relaxation properties vary across tissue regions or subjects. By addressing these limitations, MWF-MIMOSA unifies these approaches within a single multi-parametric framework. This allows for the joint estimation of MWF together with additional quantitative parameters rather than relying exclusively on one contrast mechanism or on heavily constrained compartmental assumptions.

The simulation results (Figure 3) demonstrated that the MWF-MIMOSA improved the accuracy and precision of MWF estimation compared to conventional GRE-based MWI and models utilizing fixed compartmental relaxation parameters (WM, MAE ± mean std: $Model_{MWF\text{-}MIMOSA}$, 0.008 ± 0.0080; $Model_{MGRE}$, 0.022 ± 0.0198; $Model_{Prep}$, 0.016 ± 0.0033). In-vivo validation confirmed these simulation findings. $Model_{Prep}$ tended to overestimate the MWF and showed a relatively flat MWF contrast. This is likely because its lack of $T_2^*$ and frequency-offset modeling limited its sensitivity to susceptibility- and $T_2$* decay-related tissue differences, while the fixed compartmental parameters further constrained model flexibility. As a result, regional variations in tissue properties may have been partially absorbed into the MWF estimates, leading to biased quantification and reduced contrast variation. In contrast, $Model_{MGRE}$ produced low-SNR MWF maps with high standard deviations. This reflects the ill-posed nature of estimating compartment fractions using only $T_2^*$ and frequency parameters from MGRE data (Figure 4). By comparison, $Model_{MWF\text{-}MIMOSA}$ showed the lowest standard deviation among the evaluated methods. This suggests that incorporating additional contrasts alongside joint modeling improves the estimation's conditioning and reduces its sensitivity to initialization. In addition, the complementary parameters acquired through MWF-MIMOSA allow for more comprehensive tissue characterization, which is highly relevant for investigating demyelinating, neurodevelopmental, and other WM disorders. However, the biological specificity, repeatability, and clinical utility of these parameters will require further validation.

MWF estimation was implemented using the GACELLE framework, enabling efficient joint optimization of multi-compartment parameters. Compared to conventional CPU-based, voxel-wise non-linear least-squares approaches, GACELLE processes multiple voxels in parallel on a GPU, substantially improving computational efficiency and making high-dimensional signal models highly practical. Building on this framework, we trained an MLP to surrogate the computationally intensive Bloch simulation process, which accelerated MWF estimation by more than 100-fold. The whole brain MWF estimation took approximately 20 minutes. Moreover, the MLP training loss was tailored to improve complex-valued signal fitting. As demonstrated by our ablation results (Figure S3), the proposed loss design yielded the best validation performance compared to conventional $L_2$-norm loss. Together, these technical developments make the proposed framework highly computationally efficient and well-suited for multi-parametric MW estimation.

The performance of MWF-MIMOSA was further evaluated against established MWI methods. In-vivo experiments showed that MWF-MIMOSA provides MWF maps with finer delineation of WM fiber bundles (Figure 6). Compared to MCR and GRASE, MWF-MIMOSA achieved higher spatial resolution (1 mm isotropic, versus 1.5 mm isotropic for MCR and 1x1x4 mm for GRASE) in approximately half the scan time, while simultaneously providing complementary quantitative maps. ROI analysis across five major WM regions showed good correspondence with both MCR and GRASE, with lower variability than GRASE (Figure 7). Because these methods rely on different signal mechanisms for MWF estimation, a Pearson correlation analysis was used to assess cross-method agreement. A strong correlation was observed between MWF-MIMOSA and MCR ($r$ = 0.924, $p$ = $1.15\times10^{-7}$), whereas a moderate-to-strong correlation was observed between MWF-MIMOSA and GRASE ($r$ = 0.516, $p$ = 0.034). The stronger correlation with MCR is likely due to both methods relying on $T_1$ and MGRE-based information for MWF estimation, whereas GRASE is primarily a $T_2$-based method. Other potential contributing factors include $T_2^*$ confounding effects, particularly in CST, and the relatively thick slice thickness used in GRASE (4 mm), which may increase partial-volume effects and thereby bias ROI measurements and reduce spatial specificity. In addition, the use of a tilted FOV and nonlinear registration may introduce further registration error.

Another challenge for MWI is the limited spatial resolution imposed by SNR and scan-time constraints. Integrating MZS-SSL with MWF-MIMOSA for joint reconstruction enables high-resolution multi-parametric mapping within practical scan times. The comparison of reconstruction results for 0.7 mm isotropic data obtained with MZS-SSL, SENSE (Pruessmann *et al.*, 1999), and compressed sensing (CS) (Lustig, Donoho and Pauly, 2007) were presented in Figure S1. The arrows highlight that MZS-SSL provides improved SNR than SENSE and preserves finer structural details than CS. The network was first pre-trained using data from one subject for about 18 hours, and transfer learning was then applied to each new subject. With transfer learning, the reconstruction time was reduced to approximately 4 hours per subject. This is because subject-specific self-supervised reconstruction still requires substantial fine-tuning for each new subject to adapt to subject-specific anatomy. This reconstruction time could be further reduced if a supervised network pre-trained on a large dataset were available. With a more generalizable initialization, only limited fine-tuning is needed, which could reduce the reconstruction time to about 10 min per subject (Pato Montemayor *et al.*, 2026).

Future improvements in SNR could be achieved by improving the conditioning of parameter estimation. One potential direction is the integration of diffusion-weighted imaging (DWI). Diffusion MRI is well-established for characterizing WM fiber orientation (Jellison *et al.*, 2004). Advanced diffusion models, such as neurite orientation dispersion and density imaging (NODDI) (Zhang *et al.*, 2012) and the spherical mean technique (SMT) (Kaden *et al.*, 2016), provide metrics related to neurite density and orientation dispersion. This information is directly relevant, as apparent MWF varies systematically with fiber orientation relative to the main magnetic field (Sati *et al.*, 2013; Wharton and Bowtell, 2013; Birkl *et al.*, 2021). In addition, diffusion-derived axonal volume fractions have already been combined with MGRE-based myelin imaging for whole-brain in-vivo g-ratio mapping (Jung *et al.*, 2018; Mohammadi and Callaghan, 2021). Incorporating such complementary microstructural information could help constrain the signal model, reduce the number of free parameters, and improve the robustness of MWF estimation. In the current framework, TV regularization with a fixed $\lambda$ was used to stabilize the optimization. However, a single regularization weight may not be optimal across subjects and tissue types. By providing priors on fiber orientation and compartmental volume fractions, DWI may reduce the dependence on this empirically chosen regularization, or even remove the need for it, while improving the SNR and stability of MWF estimation, particularly in regions with complex fiber architecture.

The cortical analysis revealed a pattern of heavier myelination consistent with established myeloarchitectural findings (Glasser and Van Essen, 2011; Gordon, May and Nelson, 2019; Eichert *et al.*, 2020; Plank *et al.*, 2026). Much of the existing literature relies on myelin maps derived from the ratio of $T_1$- and $T_2$-weighted images (Righart *et al.*, 2017; Lee *et al.*, 2024; Plank *et al.*, 2026). While this approach has been highly influential, $T_1$- and $T_2$-weighted contrasts are not quantitative. The resulting values are susceptible to acquisition protocols, scanner-dependent factors, transmit-field inhomogeneities (Glasser *et al.*, 2022), and non-myelin tissue properties, making the direct comparison of absolute values across studies challenging. By estimating the MWF, MWF-MIMOSA provides a more biophysically interpretable marker, as it is derived from explicit compartmental modeling of the myelin water signal rather than relative image-intensity ratios. Consequently, this framework is well-suited for studying inter-regional, inter-subject, and potentially longitudinal variations in cortical myelination. Nevertheless, further evaluation of scan-rescan repeatability remains necessary to establish the robustness and repeatability of MWF measurements obtained with MWF-MIMOSA.

By providing both $T_2$ and $T_2^*$ maps together with the tissue field map, MWF-MIMOSA enables susceptibility source separation using χ-sepnet in addition to the conventional total QSM. The diamagnetic susceptibility map has been proposed as a myelin-sensitive susceptibility contrast (Chen *et al.*, 2021; Dimov *et al.*, 2022; Li *et al.*, 2023; W. Kim *et al.*, 2023; Kan *et al.*, 2024), and therefore provides a complementary assessment of white matter tissue composition alongside MWF. However, some cerebellar fiber structures that are visible in the MWF maps are less conspicuous in the diamagnetic susceptibility maps (Figure 9). This observation does not necessarily indicate inconsistency between the two myelin-related contrasts, because MWF and diamagnetic susceptibility reflect different biophysical properties. In addition, χ-sepnet was trained using conventional 2D MESE and 3D MGRE data (Kim *et al.*, 2025), whereas the $R_2$' and field maps in this study were derived from the proposed quantitative mapping framework. This difference may affect network generalization and reduce the conspicuity of fine diamagnetic structures, particularly in anatomically complex regions such as the cerebellum. Therefore, the diamagnetic susceptibility maps should be interpreted as a complementary susceptibility-based marker of myelin rather than a direct surrogate for MWF. Further improvement in the consistency of fine white matter contrast can be achieved by fine-tuning χ-sepnet using acquired MWF-MIMOSA data.

Motion correction remains an open need, especially for high-resolution imaging and applications involving clinical and pediatric populations. Incorporating motion correction techniques, such as volumetric motion navigator (vNav) (Tisdall *et al.*, 2012, 2016; Frost *et al.*, 2016), into MWF-MIMOSA implementations would help preserve both image fidelity and quantitative reliability. Additionally, magnetisation exchange between free and macromolecular pools (Wolff and Balaban, 1989; Henkelman *et al.*, 1993; Sled and Pike, 2001) was not modeled in this study, which may have introduced bias into the estimation of myelin-related parameters (Chan, Chamberland and Marques, 2023; van der Weijden *et al.*, 2023). Incorporating Bloch-McConnell equations (Graham and Henkelman, 1997; Graf *et al.*, 2021) into the MWF-MIMOSA signal model could further improve the biophysical specificity of MWF estimation. Furthermore, because the existing literature on compartment-specific relaxometric parameters remains limited, further validation is required to fully assess the accuracy and potential applications of the estimated compartmental parameters (Figure S5). Finally, because only one patient with MS was included in this proof-of-concept, investigations in larger clinical cohorts are necessary to better understand the relationship between the quantitative maps provided by MWF-MIMOSA and underlying pathology.

# 5. Conclusion

In this study, we proposed MWF-MIMOSA for the efficient simultaneous mapping of $T_1$, $T_2$, $T_2^*$, susceptibility source separation, and MWF. By combining a multi-parametric MIMOSA acquisition with a three-compartment MWF model based on Bloch simulations, the framework enables accurate and robust myelin water estimation while providing complementary quantitative parameter maps. Integration of MZS-SSL reconstruction and the GACELLE-based MLP surrogate for parameter estimation further enabled high-fidelity whole-brain reconstruction and accelerated MWF estimation by more than 100-fold. Both simulations and in-vivo experiments demonstrated the accuracy, precision, and practical value of the proposed framework. Consequently, MWF-MIMOSA provides a promising tool for fast, high-resolution multi-parametric qMRI and may facilitate future studies of brain myeloarchitecture and related neurological disorders.

# Conflict of Interest Statement

Gian Franco Piredda and Tom Hilbert are employees of Siemens Healthineers International AG, Switzerland. The remaining authors declare no conflict of interest.

# Data Availability Statement

The data that support the findings of this study are available from the corresponding author upon reasonable request. A representative raw human imaging dataset is publicly available as an example at https://doi.org/10.5281/zenodo.19869701. The sequence and reconstruction code used in this study are available at https://github.com/yutingchen11/MWF-MIMOSA.

# Ethics Statement

This study was approved by the Institutional Review Board of Massachusetts General Hospital. All participants provided written informed consent before participation.

# Author Contributions

Yuting Chen: Conceptualization, Methodology, Software, Investigation, Writing – original draft, Writing – review & editing, Visualization; Yohan Jun: Conceptualization, Methodology, Writing – original draft, Writing – review & editing, Supervision, Funding acquisition; Hyeong-Geol Shin: Conceptualization, Methodology, Writing – original draft, Writing – review & editing; Shizhuo Li: Conceptualization, Writing – original draft, Writing – review & editing; Shohei Fujita: Conceptualization, Methodology, Writing – original draft, Writing – review & editing; Xingwang Yong: Conceptualization, Writing – original draft, Writing – review & editing, Investigation; Jiye Kim: Conceptualization, Software, Investigation, Writing – original draft, Writing – review & editing; Jongho Lee: Conceptualization, Writing – original draft, Writing – review & editing; Gian Franco Piredda: Conceptualization, Writing – original draft, Writing – review & editing; Tom Hilbert: Conceptualization, Writing – original draft, Writing – review & editing; Aneri Bhatt: Conceptualization, Writing – original draft, Writing – review & editing, Investigation; Susie Y. Huang: Conceptualization, Writing – original draft, Writing – review & editing, Resources; Huafeng Liu: Conceptualization, Writing – original draft, Writing – review & editing, Resources, Supervision, Funding acquisition; Huihui Ye: Conceptualization, Writing – original draft, Writing – review & editing; Shahin Nasr: Conceptualization, Writing – original draft, Writing – review & editing, Visualization; Borjan Gagoski: Conceptualization, Writing – original draft, Writing – review & editing, Funding acquisition; Kwok-Shing Chan: Conceptualization, Methodology, Software, Investigation, Writing – original draft, Writing – review & editing, Supervision, Funding acquisition; Berkin Bilgic: Conceptualization, Methodology, Writing – original draft, Writing – review & editing, Resources, Supervision, Funding acquisition.

# Figure legends

**Figure 1.** Overview of the MWF-MIMOSA framework. (a) Sequence diagram illustrating the acquisition modules. The sequence consists of a short TE (20 ms) $T_2$ preparation module ($T_2$ $prep_1$) followed by a FLASH readout ($RO_1$), a long TE (80 ms) $T_2$ preparation module ($T_2$ $prep_2$) followed by a FLASH readout ($RO_2$), an inversion pulse followed by two FLASH readouts ($RO_3$, $RO_4$), and a 3D multi-echo bipolar gradient echo (MGRE) module to generate $T_2^*$ and magnetic susceptibility contrasts. The dashed box details the MGRE module parameters for experiment at 1 mm isotropic resolution: TEs = [1.98, 4.56, 7.14, 9.72, 12.30, 14.88, 17.46, 20.04, 22.62, 25.20] ms and TR = 27.7 ms. Total sequence TR of MIMOSA is 6,956.5 ms. (b) Reconstruction pipeline. Undersampled multi-contrast k-space data are reconstructed using multi-contrast/-slice zero-shot self-supervised learning (MZS-SSL). The reconstruction network was implemented as an unrolled architecture with multiple units. Each unit consisted of a residual learning-based denoiser followed by a data-consistency layer (DC) solved using a conjugate gradient algorithm.

**Figure 2.** Quantitative parameter estimation pipeline of MWF-MIMOSA. (a) Single-compartment parameter estimation. Reconstructed magnitude images undergo dictionary matching to estimate $T_1$, $T_2$, and $T_2^*$ maps. Phase images are used further used for quantitative susceptibility mapping (QSM) and susceptibility source separation, incorporating an $R_2'$ map derived from $T_2$ and $T_2^*$. (b) GACELLE-based Myelin water fraction (MWF) estimation. MWF is estimated using a GACELLE-based multilayer perceptron (MLP) forward model, which iteratively updates parameters $x$ until convergence to yield the final water fraction maps.

**Figure 3.** Numerical simulation results. (a) Comparison of mean myelin water fraction (MWF) maps across 10 initializations using $Model_{Prep}$, $Model_{MGRE}$, and the proposed $Model_{MWF\text{-}MIMOSA}$ at a 45˚ fiber orientation. The corresponding error and standard deviation maps are displayed with 5x and 10x scaling, respectively. The lowest mean absolute error (MAE) and mean standard deviation (MSTD) for each tissue type are underlined. (b) Regression analysis of MWF estimation robustness using $Model_{MWF\text{-}MIMOSA}$, evaluated across MWF values ranging from 0.05 to 0.2 at parallel (0˚) and perpendicular (90˚) fiber orientations. The error bar shows the MSTD within WM.

**Figure 4.** In-vivo evaluation of myelin water fraction (MWF) estimation models. Comparison of mean MWF maps across 10 initializations for $Model_{Prep}$, $Model_{MGRE}$, and $Model_{MWF\text{-}MIMOSA}$. Rows 1 and 3 present two representative slices from a single subject. Rows 2 and 4 show the corresponding standard deviation maps across 10 initializations, scaled by a factor of 10 for visibility. For each model, the mean standard deviation (MSTD) is reported.

**Figure 5.** In-vivo multiparametric mapping. Simultaneous whole-brain estimation of $T_1$, $T_2$, $T_2^*$, quantitative susceptibility mapping (QSM), susceptibility source separation ($X_{para}$ and $X_{dia}$) and myelin water fraction (MWF) at 1 mm isotropic resolution, acquired in 4 min 55 sec using MWF-MIMOSA.

**Figure 6.** Qualitative comparison of in-vivo myelin water fraction (MWF) mapping. MWF maps from five subjects obtained using MCR, GRASE, and MWF-MIMOSA. MWF-MIMOSA achieves higher spatial resolution in a shorter acquisition time. Orange arrows highlight the improved delineation of white matter fiber bundles, particularly at the distal ends of major fiber tracts, in the MWF-MIMOSA maps.

**Figure 7.** Quantitative region of interest (ROI) analysis. **(a)** Mean and standard deviation of myelin water fraction (MWF) values across five subjects within five major white matter ROIs for MWF-MIMOSA, MCR, and GRASE. **(b)** Correlation analysis of MWF estimates across 17 white matter ROIs between MWF-MIMOSA and other methods. Each data point represents the mean MWF value from the five subjects for a specific ROI, with error bars indicating the standard deviation. Blue shading denotes stronger positive correlations.

**Figure 8.** Cortical myelin mapping. Surface visualizations of right hemisphere cortical myelin water fraction (MWF) maps for five individual subjects and the resulting group average. Regions with relatively high myelination, including the primary motor cortex (M1), primary somatosensory cortex (S1), middle temporal visual area (MT), and primary auditory cortex (A1), are indicated.

**Figure 9.** Whole-brain quantitative maps obtained using MWF-MIMOSA at 0.7 mm isotropic resolution in 9 min 48 sec. Yellow arrows indicate areas where the 0.7 mm resolution enables enhanced visualization of subtle structures, such as the internal capsule and cerebellar white matter fibers.

**Figure 10.** Clinical application in multiple sclerosis (MS). Clinical $T_2$-FLAIR images and corresponding quantitative maps obtained with MWF-MIMOSA 1 mm isotropic resolution protocol from a patient with MS. Yellow arrows highlight MS lesions, which exhibit reduced myelin water fraction (MWF) and increased intra- and extra-axonal water (IEW), and free water (FW), alongside with prolonged $T_1$, $T_2$, and $T_2^*$ relaxation times.

# Supplementary materials

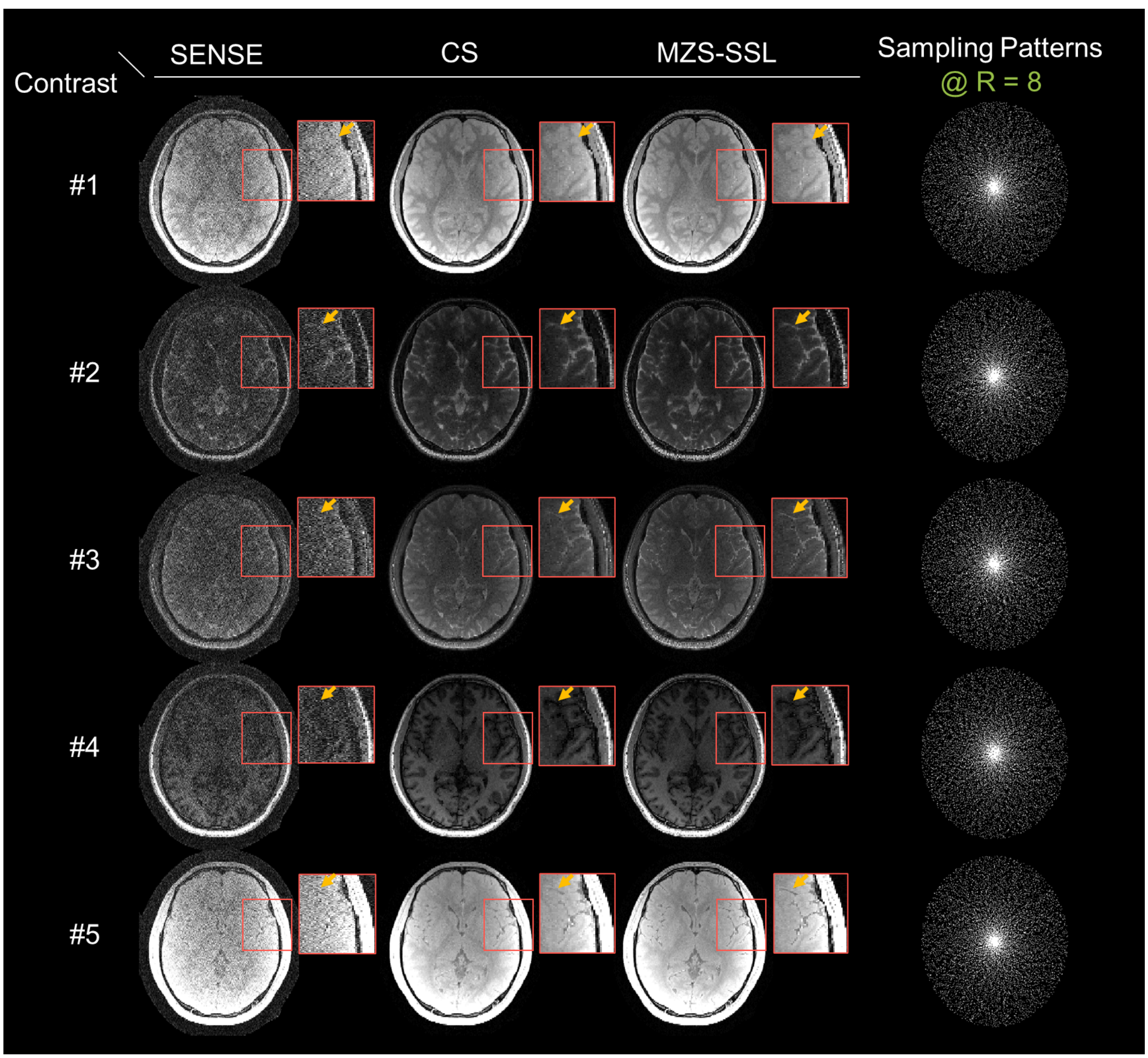

**Figure S1.** Comparison of SENSE, CS, and MZS-SSL reconstruction results for 0.7 mm isotropic data. The first five contrasts and the corresponding k-space sampling patterns used to achieve an acceleration factor (R) of 8 are shown.

# 1. MLP-Based Bloch Surrogate Model for MWF Estimation

## 1.1 Training data

Similar to the EPG-X approximation model implemented in GACELLE (Chan et al., 2025), around $10^7$ random parameter sets were generated during training with the following parameter ranges: $f_{MW} \in [0, 0.3]$, $f_{IEW} \in [0, 1]$, $T_{1,MW} \in [100, 400]$ ms, $T_{1,IEW} \in [700, 1500]$ ms, $R_{2,MW} \in [25, 100]$ Hz, $R_{2,IEW} \in [10, 17]$ Hz, $R^*_{2,MW} \in [50, 200]$ Hz, $R^*_{2,IEW} \in [2, 50]$ Hz, $\triangle\omega_{\mathrm{MW}} \in [-0.05, 0.25]$ ppm, $\triangle\omega_{\mathrm{IEW}} \in [-0.1, 0.05]$ ppm, $\triangle\omega_{\mathrm{IEW}} \in [-0.1, 0.05]$ ppm, and $B_1^+ \in [0.65, 1.35]$. For each mini-batch voxel, $f_{MW}$ and $f_{IEW}$ were randomly sampled under the constraint $f_{MW} + f_{IEW} \leq 1$ and the $f_{FW}$ was set as 1- $f_{MW}$ - $f_{IEW}$. All FW relaxation constants were fixed in this implementation.

The MLP network contains seven hidden layers of filter size [80 120 160 180 240 260 300], followed by leaky RELU activation functions. The network takes 16-channel physics-informed input features described in the GitHub repository and outputs two channels corresponding to the predicted real and imaginary signal components.

Training labels were generated on the fly using the Bloch simulations based on the three-compartment MWF-MIMOSA signal model. The input parameters were transformed to 16 physics-informed feature vectors to explicitly encode tissue-related physical priors. The features comprise five categories: (1) tissue-composition features, including $f_{MW}$ and $f_{IEW}$; (2) relaxation features, including $R^*_{2,MW/IEW}$-dependent decay terms encoding echo-time information, as well as normalized $T_{1,MW/IEW}$ and $R_{2,MW/IEW}$; (3) frequency features, including $\triangle\omega_{\mathrm{MW/EW}}$; and (4) transmit field inhomogeneity features, represented by the $B_1^+$ scaling factor; (5) temporal features, including a normalized acquisition index and a one-hot code indicating whether the acquisition originated from a FLASH readout or the MGRE readout. All continuous features were normalized to reduce scale disparities across inputs and improve training stability.

## 1.2 Loss function

To avoid the suboptimal magnitude-phase behavior of conventional complex-domain losses, a weighted relative complex loss (WRCL) was used to train the MLP to approximate Bloch-simulated complex signals. Similar to previous work (Xue et al., 2025; Terpstra et al., 2022; Barron, 2019), a continuously differentiable operator for the complex signals was defined as:

$$\|z\|_\epsilon = \sqrt{Re(z)^2 + Im(z)^2 + \epsilon},$$

where $\epsilon$ is a small constant introduced for numerical stabilization and stable backpropagation. Unlike the standard $L_2$ norm, the inclusion of $\epsilon$ avoids non-smooth behavior near zero magnitude and stabilizes optimization in low-signal regimes.

Let $S_n$ and $\hat{S}_n$ denote the ground-truth and predicted complex signals, respectively, for the *n*-th sample in a mini-batch of size *N*. The complex prediction error was defined as:

$$E_n = \|\hat{S}_n - S_n\|_\epsilon,$$

and the reference magnitude was defined as:

$$M_n = \|S_n\|_\epsilon.$$

Because multi-compartment models inherently involve markedly different signal fractions and decay rates, a soft magnitude-dependent weighting term was further applied:

$$w_n = \frac{M_n^2}{M_n^2 + \tau^2},$$

where $\tau$ controls the degree of suppression for low-amplitude samples.

The final training objective was then given by:

$$\mathcal{L}_{MLP} = \frac{1}{N}\sum_{n=1}^{N} w_n \left(\frac{E_n}{M_n}\right).$$

This formulation combines relative error normalization and soft magnitude-aware weighting, so that optimization is less dominated by high-amplitude samples while remaining stable in fully relaxed or near-zero signal regimes.

## 1.3 Implementations

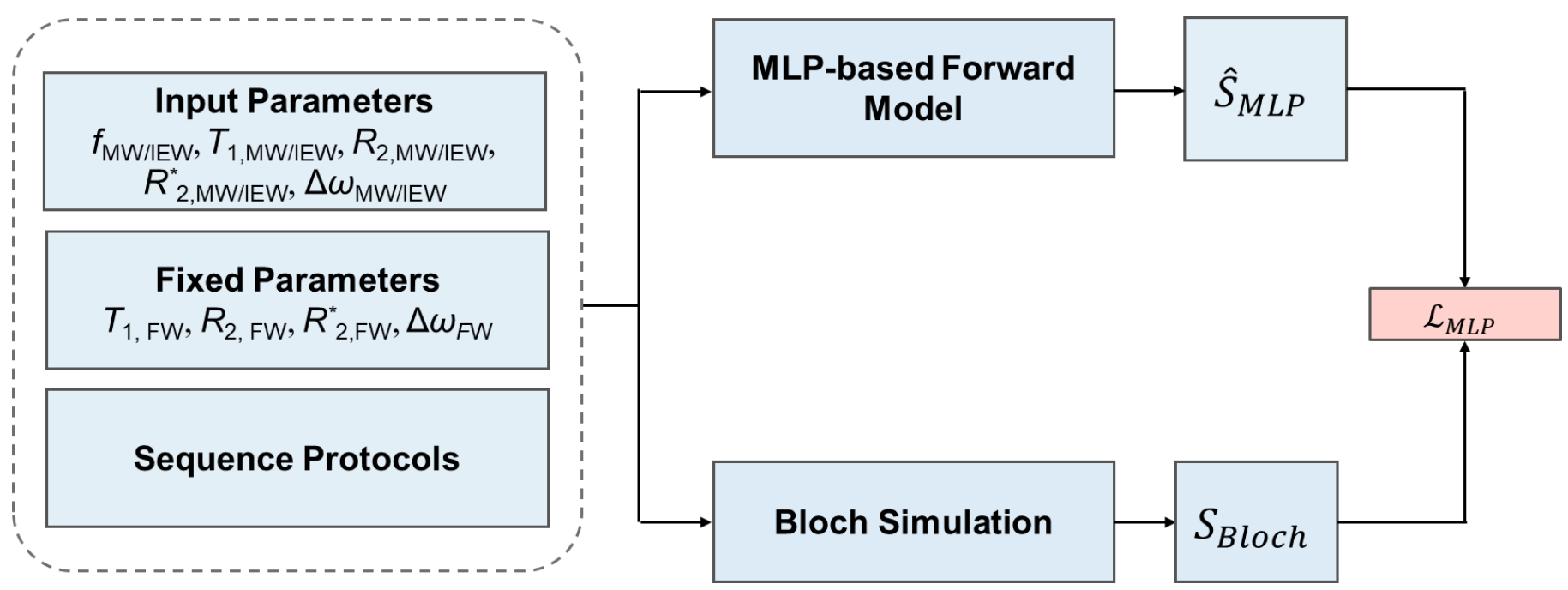

**Figure S2.** MLP training pipeline. During training, input parameters were randomly sampled within predefined ranges and transformed into 16 physics-informed feature vectors based on the fixed relaxometric parameters of the free-water compartment and the sequence protocol. The MLP-based forward model takes the 16-channel feature vector as input and outputs a two-channel signal, $S_{MLP}$, corresponding to the predicted real and imaginary components. The training labels, $S_{Bloch}$, were generated using Bloch simulations based on the three-compartment MWF-MIMOSA signal model. The training loss, $L_{MLP}$, was used to optimize the network through backpropagation.

Training was performed in MATLAB (MathWorks, Natick, MA, USA) on an NVIDIA RTX6000 GPU (NVIDIA, Inc., Santa Clara, CA, USA) for 1000 epochs. Model parameters were optimized using the Adam optimizer with an initial learning rate of 0.01 and a decay rate of $1\times10^{-4}$ per iteration. The total training time was about approximately 4 hours.

Validation was performed using an in-distribution synthetic test set comprising $2\text{x}10^{4}$ randomly generated parameter combinations, sampled from the same parameter ranges. The trained network predicted the complex-valued signals and the predictions were compared against the Bloch-simulated reference signals. Performance was quantified using complex normalized root-mean-square error (NRMSE).

1.4 Ablation study

We first conducted an ablation study to investigate the role of temporal features in the proposed MLP framework. As shown in Figure S2 (a), removing temporal features led to a marked increase in validation complex NRMSE to 43.13%. Using only a one-hot temporal code reduced the error to 5.77%, whereas using only a normalized acquisition index further reduced it to 0.26%. The lowest validation complex NRMSE was achieved when both temporal features were included (0.17%). These results indicate that temporal features are particularly important, as they enable the network to distinguish signals acquired from different sequence modules.

To assess the contribution of the two key design choices in WRCL, we performed an ablation study on the weighting and relative normalization terms. Specifically, we compared four variants: a conventional $L_2$ norm loss, a plain complex loss without weighting or relative normalization (CL), a weighted-only version (WCL), a relative-only version (RCL), and the full WRCL. This design allowed us to isolate the effects of soft magnitude-aware weighting and relative error normalization on training stability and signal fidelity across compartments with substantially different amplitudes and decay characteristics. As illustrated in Figure S2 (b), WRCL achieved the lowest complex NRMSE (0.17%) among all compared losses, outperforming WCL, CL, RCL, and $L_2$. Using the weighting term alone (WCL) provided a modest improvement over the plain complex loss (CL), suggesting that soft magnitude-aware weighting helped balance samples with different signal amplitudes. In contrast, using relative normalization alone (RCL) resulted in worse performance than CL, indicating that normalization by the reference magnitude, although intended to reduce scale imbalance, could make optimization less stable when applied without additional control. When combined, however, the weighting and relative terms acted complementarily, and the full WRCL yielded the best overall accuracy. Notably, all complex-domain variants substantially outperformed the conventional $L_2$ loss, further supporting the benefit of explicitly accounting for complex-valued signal behavior in this task.

The effects of the weighting hyperparameter $\tau$ and the smoothing constant $\varepsilon$ were further evaluated to balance numerical stability and accurate complex-signal fitting. As shown in Fig S2 (b), the best performance was obtained at $\tau = 10^{-2}$ and $\varepsilon = 10^{-4}$. Overall, moderate $\tau$ values gave lower complex NRMSE than larger $\tau$, indicating that mild suppression of low-amplitude samples was helpful, whereas overly strong weighting was detrimental. In addition, a large smoothing constant ($\varepsilon = 10^{-3}$) consistently degraded performance, suggesting that excessive smoothing weakened the fidelity of complex-signal matching. This trend is consistent with the design of the loss: $\tau$ controls the down-weighting of low-amplitude samples, whereas $\varepsilon$ controls the smoothness of the complex modulus; overly aggressive settings for either term may reduce sensitivity to weak but informative signal components.

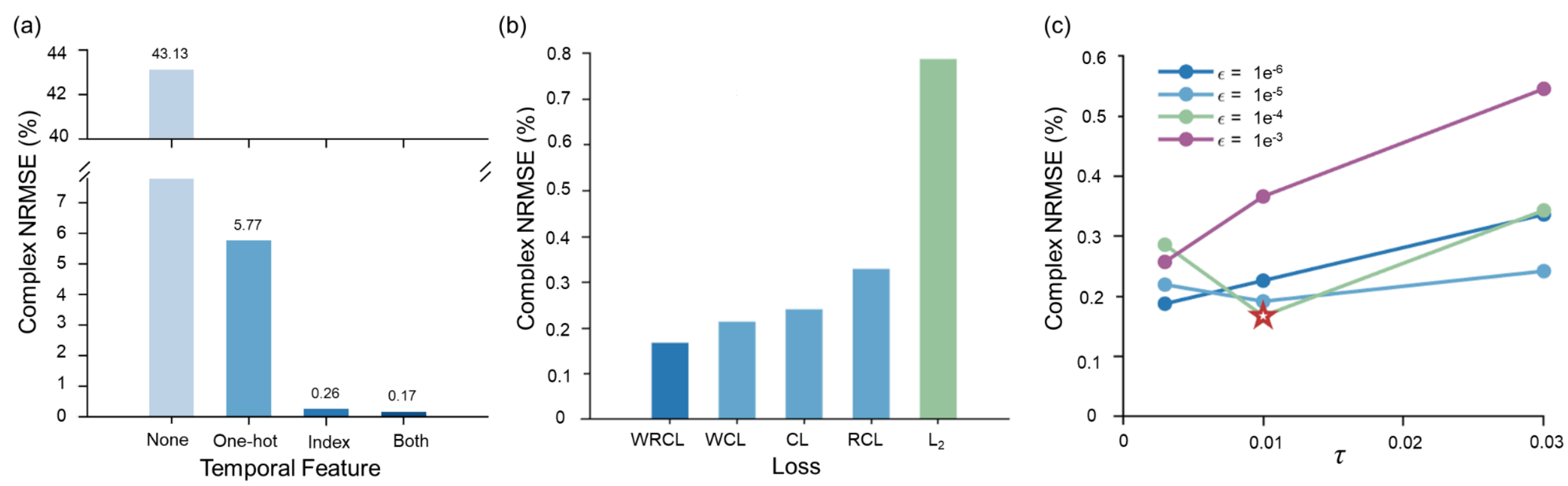

**Figure S3.** Ablation study results evaluated by validation NRMSE. (a) Effect of temporal feature design, including no temporal feature (None), a one-hot temporal code only (One-hot), an acquisition index only (Index), and the full temporal feature set (Both). (b) Comparison of different training losses, including a conventional $L_2$ norm loss, a plain complex loss without weighting or relative normalization (CL), a weighted-only version (WCL), a relative-only version (RCL), and the full weighted relative complex loss (WRCL). (c) Effect of different choices of the weighting hyperparameter $\tau$ and the smoothing constant $\varepsilon$ in the proposed WRCL loss. The hyperparameter setting with lowest NRMSE is marked with a star.

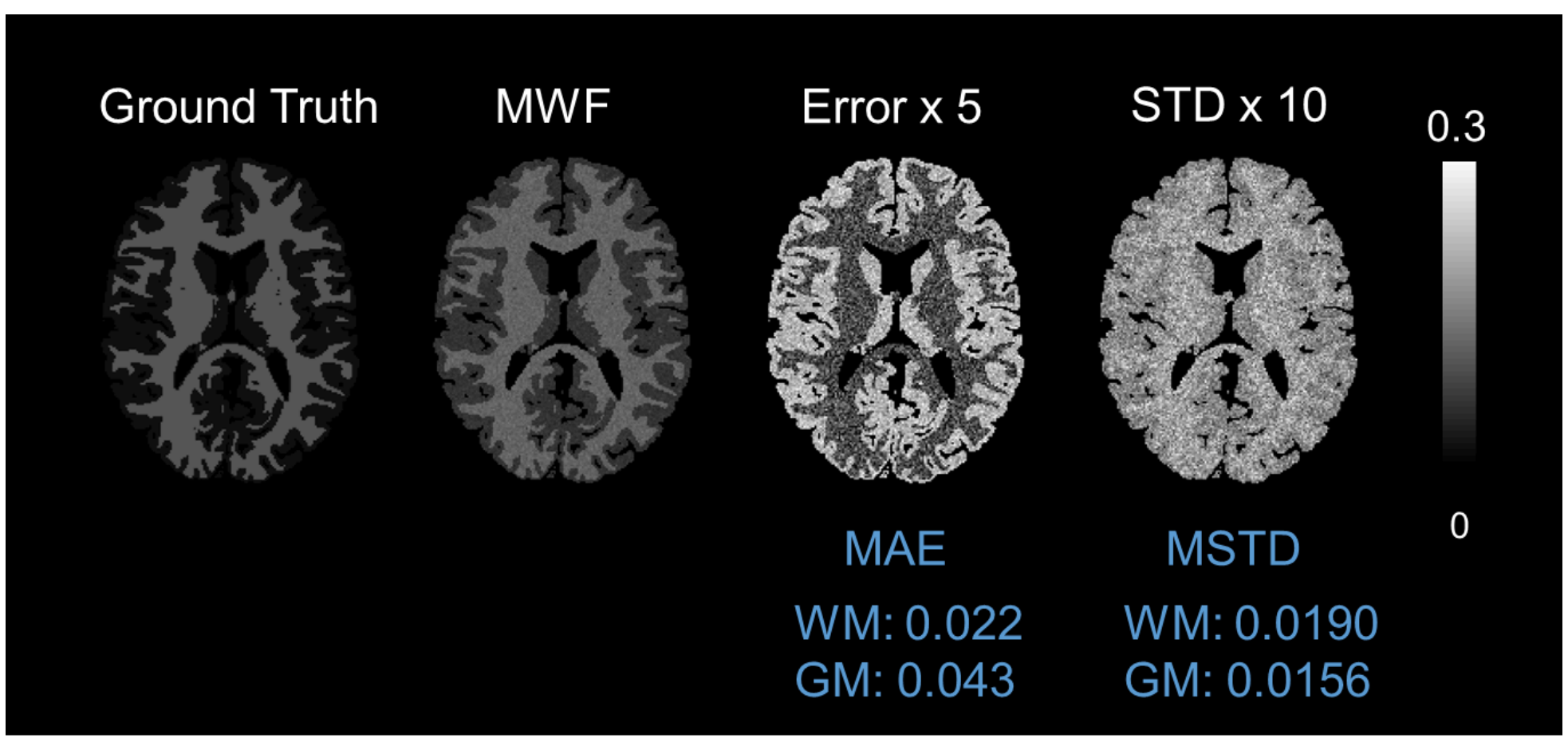

**Figure S4.** Numerical simulation results of $Model_{Prep}$, with the fixed compartmental $T_{1,MW/IEW}$ and $T_{2,MW/IEW}$ values randomly sampled from [100, 200] ms, [900, 1100] ms, [10, 20] ms, and [60, 80] ms, respectively. The error map and standard deviation map are shown at x5 and x10 scale, respectively.

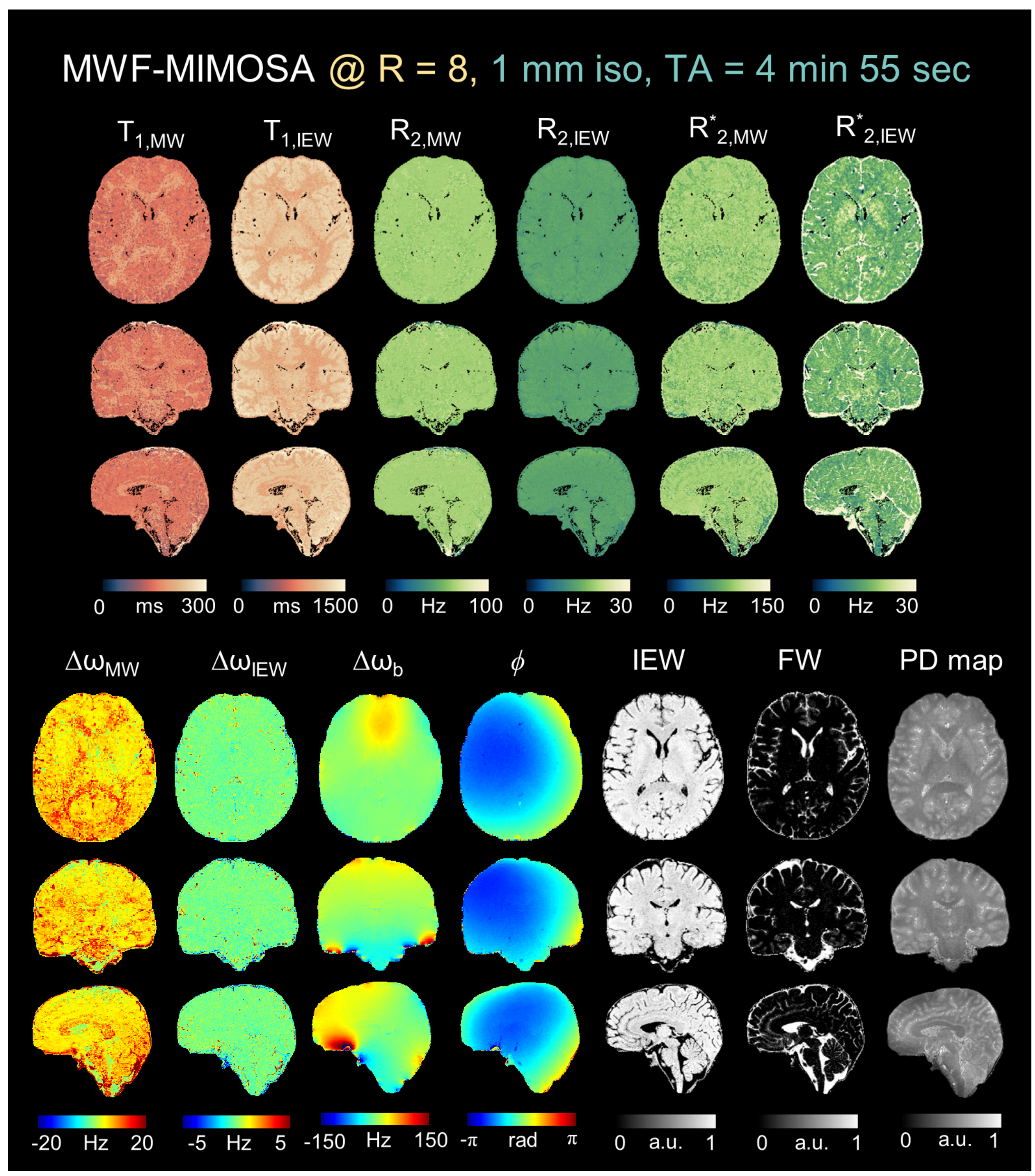

**Figure S5.** Additional estimated compartmental parameter maps, including $T_1$, $R_2$, $R_2^*$, frequency-offset, and fraction maps of the IEW and FW pools, together with the background frequency-offset $\Delta\omega_b$, initial phase $\phi$, and PD maps.

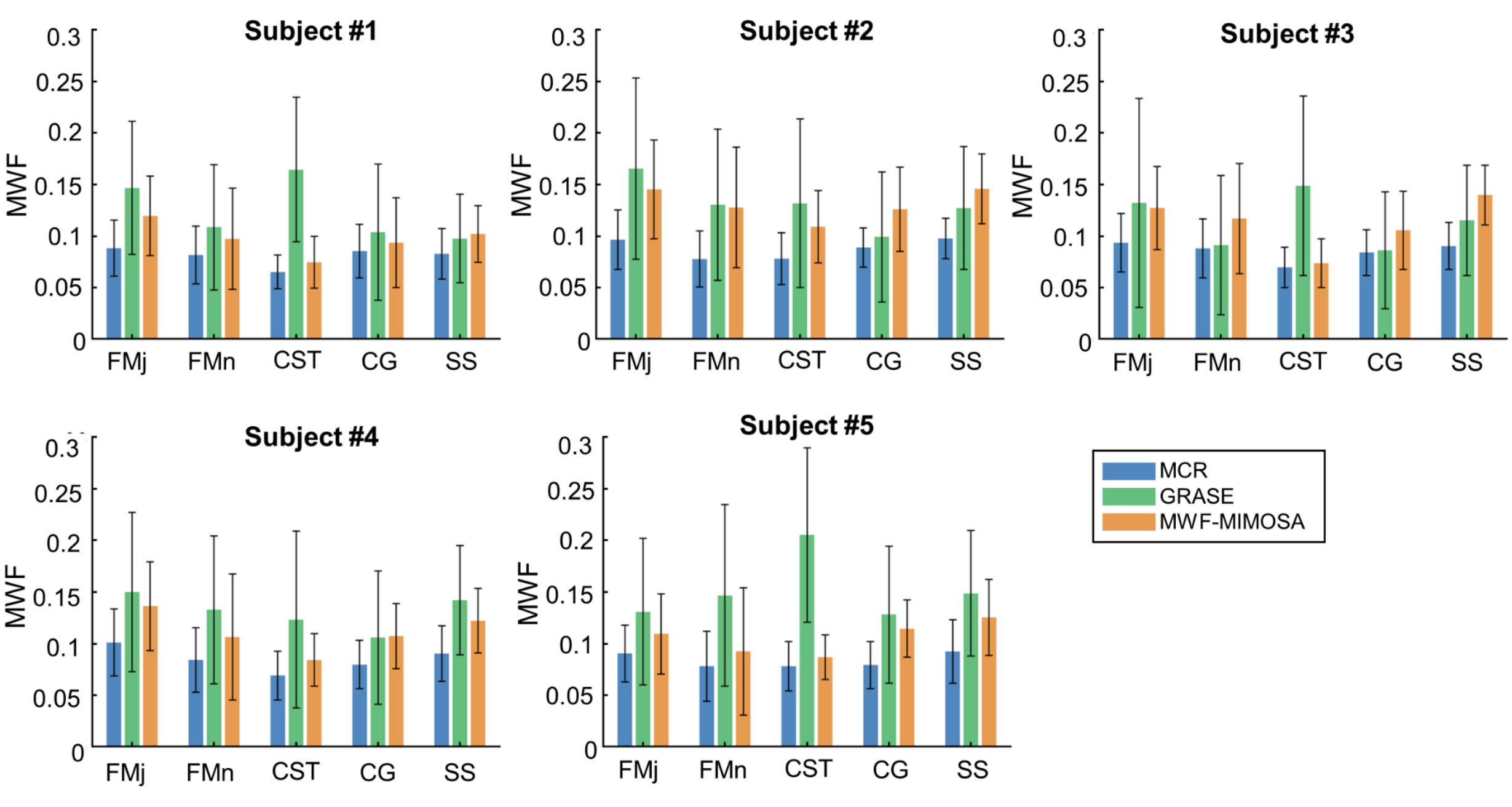

**Figure S6.** Subject-wise ROI analysis of MWF for MIMOSA, MCR, and GRASE across five white-matter ROIs, showing the mean and standard deviation of MWF.